\documentclass[12pt]{article}
\usepackage{graphicx}
\usepackage{amsfonts} 
\usepackage{helvet}        
\usepackage{courier}        
\usepackage{amsmath,bm}
\usepackage{amsthm}
\usepackage{amssymb}
\usepackage{subfigure}
\usepackage{blindtext}
\usepackage{url}
\usepackage{hyperref}
\usepackage{apacite}
\usepackage{dsfont}
\usepackage{comment}
\usepackage{natbib}
\usepackage{type1cm}        
\usepackage[toc,page]{appendix}
\usepackage{makeidx}         
\usepackage{multicol}        
\usepackage{graphicx}      
\usepackage[bottom]{footmisc}
\usepackage{colortbl,booktabs}
\usepackage{tablefootnote}
\usepackage{hyperref}
\usepackage[dvipsnames]{xcolor}
\usepackage{tikz}
 
\usepackage{algorithm}
\usepackage{algpseudocode}
 
\usepackage{authblk}
\usepackage{amsthm}

\theoremstyle{definition}
\newtheorem{definition}{Definition}

\theoremstyle{plain}

\newcommand{\keywords}[1]{%
  \noindent\textbf{Keywords:} #1
}
 
\usepackage{setspace}
\usepackage[top=2.5cm, bottom=2.5cm, left=0.7in, right=0.7in, includefoot]{geometry}

\title{Generalized Infinite Hidden Markov Models} 
 
\author[1]{Giuseppe Mignemi} 
\author[2]{Daniele Tancini}
\author[2]{Francesco Bartolucci}
 
\affil[1]{Institute of Computer Science - Czech Academy of Science}
\affil[2]{Department of Economics - University of Perugia}

\date{}

\begin{document}

 \maketitle
 
\begin{abstract}
We introduce a flexible Bayesian nonparametric hidden Markov model for longitudinal data based on the hidden hierarchical Dirichlet process. Unlike the classical infinite hidden Markov model, the proposed model allows transition probabilities to change over time. In particular, this model provides a framework that operates between homogeneous and heterogeneous  Markov model formulations, allowing the rows of the transition probability matrix to be clustered according to their temporal evolution. Inference is performed through a Markov chain Monte Carlo algorithm, which enables the joint estimation of model parameters, latent variables, and number of states, within a unified framework. Simulation studies under both homogeneous and heterogeneous transition regimes show that the proposed model tends to recover the true latent structure, whereas the infinite hidden Markov model tends to overestimate the number of states when transitions vary over time. Two applications illustrate the ability of the model to detect changes in the transition dynamics.

\end{abstract}

 \small
 \keywords{Bayesian nonparametrics; Dirichlet process; Latent variable models; Markov chain Monte Carlo}
 
 \normalsize

\section{Introduction}
In the analysis of longitudinal data, accounting for unobserved heterogeneity in a flexible and interpretable way is crucial. Hidden Markov models (HMMs) achieve this by assuming that the observed process is driven by a discrete latent process following a Markov chain, with transition probabilities that fully encode the temporal dependence structure; see \cite{bartolucci2012latent} and \cite{zucchini2009hidden} for comprehensive reviews. When the number of latent states of an HMM is unknown, a Bayesian nonparametric approach based on the hierarchical Dirichlet process (HDP, \citealp{Teh_2006}) provides a natural infinite-state extension, known as the HDP-HMM or infinite HMM \citep{Beal_2002, Fox_2008}. In this model, each row of the infinite transition matrix is a random measure distributed as a Dirichlet process (DP), and the rows are tied together through a shared DP-distributed base measure. This hierarchical construction guarantees that the same set of destination states can be revisited across different departure states, while imposing time-homogeneity. 

It is worth noting that many longitudinal and time-series applications call for time-heterogeneous transition mechanisms, which can change with time while still borrowing strength across time points and across states \citep{hughes1999non,bartolucci2007class}. Accommodating such a heterogeneity requires grouping the transition distributions themselves across time points; a natural nonparametric prior for clustering distributions, in addition to the observations they generate, is the nested Dirichlet process (NDP, \citealp{Rodriguez_2008}). However, as shown by \cite{camerlenghi2019latent}, the NDP suffers from degeneracy. In fact, as soon as two samples share a single observation, their underlying random distributions are forced to coincide almost surely. In the longitudinal setting, this means that the transition kernels at two different time points collapse to the very same distribution as soon as a single common destination state is observed, which is clearly undesirable.

The hidden hierarchical Dirichlet process (HHDP) introduced by \citet{Lijoi_2023} resolves the degeneracy at issue by blending the HDP and the NDP. This approach retains the ability of the NDP to cluster distributions avoiding its degeneracy through an HDP-distributed layer that accommodates shared atoms across different distributions, and hence borrow information, exactly as in the HDP. 

Building on the HHDP, we propose the generalized infinite hidden Markov model (GIHMM). In this construction the transition kernel associated with departure state 
is allowed to be time-varying. More in detail, the proposed GIHMM provides a flexible framework that lies between the homogeneous and heterogeneous HMM formulations. In particular, the model allows the rows of the transition probability matrix to be clustered in a suitable way. 

Our proposal offers several advantages. First, it jointly accounts for heterogeneity at two levels. At the state level, units are clustered into latent states at each time point. At the distributional level, the transition distributions of each departure state are clustered across time, so that changes in the transition dynamics can be detected. Both levels are jointly accommodated within a single framework. Second, the model offers a compromise between flexibility and parsimony. Homogeneous HMMs assume a single transition matrix for the whole process, which may be too restrictive in the presence of change points. Fully heterogeneous HMMs, on the other hand, specify a different transition matrix at each time point, thus limiting the borrowing of information across time points even when the data suggest a 
homogeneous transition dynamic. The proposed model bridges these two extremes and includes both as special cases: each row of the transition matrix may change over time, but the same transition distribution can be shared across several time points. The temporal structure of the transition dynamics is informed by the data rather than a priori imposed. Third, all model components can be jointly estimated within a single Markov chain Monte Carlo (MCMC) algorithm, avoiding the comparison of a potentially prohibitive number of parametric specifications. In particular, we consider a Metropolis-within-Gibbs algorithm, which is based on the finite approximation of the DP \citep{muliere1998approximating,ishwaran2001gibbs}. However, instead of truncating the stick-breaking representation of the DP, we use the finite Dirichlet approximation of \cite{ishwaran2002exact}.

The proposed model and the corresponding MCMC algorithm are applied to both synthetic and real datasets. On generated data, we illustrate the competitiveness of our approach with respect to the standard infinite HMM, showing that it accurately recovers both heterogeneous and homogeneous transition structures. In the real data applications, the model reveals temporal dynamics in the transition mechanism that would remain undetected under standard homogeneous formulations.

The remainder of the paper is organized as follows. Section \ref{sec:2} provides a review of Bayesian nonparametric priors. Section \ref{sec:ext} considers a multivariate version of the HHDP and introduces the proposed GIHMM. Section \ref{sec:4} describes the MCMC algorithm developed for posterior inference and discusses relevant computational aspects of the proposed approach. Section \ref{sec:5} compares the GIHMM with the standard infinite HMM using synthetic data and demonstrates the advantages of the GIHMM through applications to two economic panel datasets. Limits and future direction are discussed in Section \ref{sec:6}. 

\section{Preliminaries on Bayesian Nonparametric Priors}\label{sec:2}
In this section, we provide a brief overview of Bayesian nonparametric priors. In particular, we first introduce the DP and then we discuss some extensions, that are, the HDP, the NDP, and the HHDP.

Let $\mathbb{X}$ be a complete and separable metric space and let $P_0$ be a nonatomic probability measure on $\mathbb{X}$. A random probability measure $G$  
follows a DP \citep{ferguson1973bayesian}, that is, $G\sim DP(\alpha,P_0)$, if, for every finite measurable partition $\{A_1,\dots,A_m\}$ of $\mathbb{X}$,
$$
\bigl(G(A_1),\dots,G(A_m)\bigr)\sim \mathcal{D}ir\bigl(\alpha P_0(A_1),\dots,\alpha P_0(A_m)\bigr),
$$
where $\alpha>0$ is the concentration parameter, $P_0$ is the base measure, and $\mathcal{D}ir(\cdot)$ denotes the Dirichlet distribution. The base measure $P_0$ is the \textit{prior guess} of $G$, in the sense that $\mathbb{E}[G(A)]=P_0(A)$ for every measurable $A\subseteq\mathbb{X}$, while $\alpha$ controls the concentration of $G$ around $P_0$. \citet{sethuraman1994constructive} showed that $G\sim DP(\alpha,P_0)$ admits, almost surely, the stick-breaking representation
\[
G=\sum_{k\geq1}\pi_k\,\delta_{X_k}, \qquad \pi_k=V_k\prod_{l=1}^{k-1}(1-V_l), \qquad V_k\overset{i.i.d.}{\sim}\mathcal{B}eta(1,\alpha), \qquad X_k\overset{i.i.d.}{\sim}P_0,
\]
where $\delta$ denotes the dirac measure.  
Hence $G$ is almost surely discrete, and the law of the weights $\{\pi_k\}_{k\geq1}$ is the Griffiths--Engen--McCloskey distribution, denoted by $GEM(\alpha)$ \citep{Ewens_1990}. Since $P_0$ is nonatomic, the atoms $X_k$ are almost surely distinct, so that ties among observations sampled from $G$ arise only from its discreteness.

An extension of the previous DP is the HDP, which has been studied from a theoretical point of view in \cite{Teh_2006}. An HDP can be defined considering
$$G_0 \sim DP(\gamma, H),$$
where $G_0$ is usually called global measure, while $\gamma > 0$ is the concentration parameter, and $H$ is the base measure. A vector of random probability measures $(G_1,\dots,G_C)$ on $\mathbb{X}$ is a hierarchical Dirichlet process, written $(G_1,\dots,G_C)\sim HDP(\beta,\beta_0;H)$, if
$$
G_c\mid G_0\overset{i.i.d.}{\sim}DP(\beta;G_0), \quad c=1,\dots,C, 
$$
and
$$
G_0\sim DP(\beta_0;H),
$$
where $H$ is a nonatomic probability measure on $\mathbb{X}$, $\beta_0>0$ is the concentration parameter of the global measure $G_0$, and $\beta>0$ controls the concentration of each $G_c$ around $G_0$. Since $G_0$ is almost surely discrete, the population-specific measures $G_c$ are supported on the same countable set of atoms, those of $G_0$, and differ only in the weights assigned to them. This induces clusters of observations that are shared across the $C$ populations, which is the mechanism exploited by the infinite HMM to allow the same destination states to be reached from different departure states; see \cite{Teh_2006} for a complete review. Furthermore, it is important to
note that the model is characterized by $\mathbb{P}(G_c = G_{c'}) = 0$
for any $c \neq c'$; see \cite{Camerlenghi_2019} and \cite{argiento2020hierarchical} for additional comments. This means that the HDP cannot be used to cluster, for example, both populations and observations. To achieve this, one has to rely on priors induced by nested structures, the most popular being the NDP.

The NDP, introduced by \cite{Rodriguez_2008}, is the most widely used nonparametric prior allowing to cluster both observations and populations. However, as shown in \cite{camerlenghi2019latent} and already mentioned above, it suffers from a degeneracy issue, 
because even a single tie shared across samples is enough to group the $C$ population distributions into a single cluster. \cite{Lijoi_2023} propose a model that combines the HDP and the NDP in a way to use their strengths, with the goal of achieving a more flexible clustering method. In particular, the authors investigate a nonparametric prior and derive a closed-form expression for the induced distribution of the random partition. According to \cite{Lijoi_2023}, the HHDP may be defined as follows
$$\{G_{1}, \ldots, G_{C}\} \sim HHDP(\alpha, \beta, \beta_0, H),$$ 
where
\begin{gather}
G_{c} \lvert Q \overset{\text{i.i.d.}}{\sim} Q, \quad Q = \sum_{k \geq 1} \pi_{k} \delta_{G_{k}^*}, \quad \{\pi_{k}\}_{k \geq 1} \sim GEM(\alpha), \quad \{G_{k}^*\}_{k \geq 1} \sim HDP(\beta, \beta_0;H), \label{HHDP_def}
\end{gather}
with $\{\pi_{k}\}_{k \geq 1}$ and $\{G_{k}^*\}_{k \geq 1}$ independent. 
In Equation \eqref{HHDP_def}, $GEM(\alpha)$ governs how the atoms $\{G_{k}^*\}_{k\geq1}$ are mixed into $Q$. In particular, a small value of $\alpha$ corresponds to most of the mass concentrated on a few atoms, whereas as $\alpha$ increases the mass thinly spreads over many atoms, making such ties increasingly unlikely. Analogously, the weights of $G_k^*$ and $G_0$ follow $GEM(\beta)$ and $GEM(\beta_0)$, respectively, with the same interpretation at the lower levels of the hierarchy.
Assuming $C \rightarrow \infty$, the sequence $\{G_{k}^*\}_{k\geq1}$ can be more conveniently represented as
\begin{gather*}
G_{k}^*=\sum_{l\geq 1} \omega_{k,l}\; \delta_{Z_{k,l}} \quad Z_{k,l}|G_0 \overset{i.i.d.}{\sim} G_0, \quad G_0 = \sum_{l \geq 1}\omega_{0,l}\; \delta_{X^*_l}, \quad X^*_l \overset{i.i.d.}{\sim} H, \\
\{\omega_{k,l}\}_{l \geq 1} \overset{i.i.d.}{\sim} GEM(\beta), \quad \{\omega_{0,l}\}_{l \geq 1} \sim GEM(\beta_0).
\end{gather*}
In this case independence is assumed between the sequences $\{\omega_{k,l}\}_{l \geq 1}$ and $\{Z_{k,l}\}_{l \geq 1}$, and between $\{\omega_{0,l}\}_{l \geq 1}$ and $\{X_l^*\}_{l \geq 1}$. Leveraging the stick-breaking representation and the closure property of the DP with respect to grouping, each $G_{k}^*$ is a re-weighted sum of the atoms of the base measure $G_0$:
\[
G_{k}^* = \sum_{l \geq 1} \omega^*_{k,l} \; \delta_{X^*_l} \quad \text{and} \quad G_0 = \sum_{l \geq 1} \omega_{0,l} \; \delta_{X^*_l},
\]
where $\{\omega^*_{k,l}\}_{l \geq 1} \lvert \boldsymbol{\omega}_0 \overset{i.i.d.}{\sim} DP(\beta;\boldsymbol{\omega}_0)$, $\boldsymbol{\omega}_0 = \{\omega_{0,l}\}_{l \geq 1} \sim GEM(\beta_0)$, and $X^*_l\overset{i.i.d.}{\sim}H$ for $l \geq 1$. The HHDP previously described presents a natural application in situations where a collection of data are recorded under different experimental conditions, and so we get different samples.

\section{Multidimensional HHDP and Generalized Infinite HMM}\label{sec:ext}

In this section we describe an extension of the HHDP proposed by \cite{Lijoi_2023}, which we refer to as multidimensional HHDP. Then, we show how to use the proposed multidimensional HHDP for a GIHMM.

\subsection{Multidimensional HHDP}\label{sec:extmod}

A further extension of the model introduced by \cite{Lijoi_2023} is the following. Let us define by $\mathcal{S} \subseteq \mathbb{N}$, and let us also assume that a time index $t=2,\ldots,T$, with $T \in \mathbb{N}$, is observed. 
\begin{definition}\label{def:mhhdp}
The random probability measures $\{G_{j,2},\dots,G_{j,T}\}_{j\in\mathcal{S}}$ represent a multidimensional hidden hierarchical Dirichlet process (MHHDP) if
\begin{equation}
G_{j,t} \mid Q_j \overset{\text{i.i.d.}}{\sim} Q_j, \quad Q_j = \sum_{k \geq 1} \pi_{k,j} \delta_{G_{k,j}^*}, \quad \{\pi_{k,j}\}_{k \geq 1} \overset{\text{i.i.d.}}{\sim} GEM(\alpha), \label{GIHMM_def1}
\end{equation}
\begin{equation}
G_{k,j}^* \mid G_0 \overset{\text{i.i.d.}}{\sim} DP(\beta; G_0), \quad G_0 \sim DP(\beta_0; H), \label{GIHMM}
\end{equation}
for $k\geq1$ and $j\in\mathcal{S}$, with $\{\pi_{k,j}\}_{k \geq 1}$ i.i.d.\ across $j\in\mathcal{S}$ and independent of $\{G_{k,j}^*\}_{k\geq1,\,j\in\mathcal{S}}$ and $G_0$. In the sequel we write $\{G_{j,2},\dots,G_{j,T}\}_{j\in\mathcal{S}} \sim MHHDP(\alpha,\beta,\beta_0;H)$.
\end{definition}
Note that, time is indexed by $t=2,\ldots,T$, which is convenient for the formulation of the proposed GIHMM introduced in the next section. However, the same MHHDP can also be formulated by indexing time as $t=1,\ldots,T$. In addition, notice that, the i.i.d.\ assumption on the $G_{k,j}^*$ holds across both $k\geq1$ and $j\in\mathcal{S}$, so that all departure states share the same base measure $G_0$. Hence, for each $j\in\mathcal{S}$, the collection $\{G_{j,2},\dots,G_{j,T}\}$ is an $HHDP(\alpha,\beta,\beta_0;H)$ as in \cite{Lijoi_2023}, while the collection $\{G_{k,j}^* : k\geq1,\ j\in\mathcal{S}\}$ is an $HDP(\beta,\beta_0;H)$, so that the same destination states can be reached from different departure states, exactly as in the infinite HMM. In the limiting case $\alpha\to 0$ each state $j$ uses the same transition distribution at all time points and the infinite HMM is recovered, whereas for $\alpha\to\infty$ no ties occur among time points and $\{G_{j,t}: j\in\mathcal{S},\ t=2,\dots,T\}$ is an HDP with populations indexed by the pairs $(j,t)$. As in \cite{Lijoi_2023}, the infinite hidden sequence $\{G_{k,j}^*\}_{k\geq1}$ can be represented as
\begin{gather*}
G_{k,j}^*=\sum_{l\geq 1} \omega_{k,j,l}\; \delta_{W_{k,j,l}} \quad W_{k,j,l}|G_0 \overset{i.i.d.}{\sim} G_0, \quad G_0 = \sum_{l \geq 1}\omega_{0,l}\; \delta_{X^*_l}, \quad X^*_l \overset{i.i.d.}{\sim} H, \\
\{\omega_{k,j,l}\}_{l \geq 1} \overset{i.i.d.}{\sim} GEM(\beta), \quad \{\omega_{0,l}\}_{l \geq 1} \sim GEM(\beta_0).
\end{gather*}
In this case independence is assumed between the sequences $\{\omega_{k,j,l}\}_{l \geq 1}$ and $\{W_{k,j,l}\}_{l \geq 1}$, and between $\{\omega_{0,l}\}_{l \geq 1}$ and $\{X_l^*\}_{l \geq 1}$. Using the stick-breaking representation and the closure property of the DP, each $G_{k,j}^*$ is a re-weighted sum of the atoms of $G_0$, that is, 
\[
G_{k,j}^* = \sum_{l \geq 1} \omega^*_{k,j,l} \; \delta_{X^*_l},  \quad G_0 = \sum_{l \geq 1} \omega_{0,l} \; \delta_{X^*_l},
\]
where $\{\omega^*_{k,j,l}\}_{l \geq 1} \lvert \boldsymbol{\omega}_0 \overset{i.i.d.}{\sim} DP(\beta;\boldsymbol{\omega}_0)$, $\boldsymbol{\omega}_0 = \{\omega_{0,l}\}_{l \geq 1} \sim GEM(\beta_0)$, and $X^*_l \overset{i.i.d.}{\sim}H$ for $l \geq 1$. As in Section \ref{sec:2}, $H$ is a nonatomic probability measure on $\mathbb{X}$, so that the atoms $\{X^*_l\}_{l\geq1}$ are almost surely distinct. 

\begin{figure}[!htbp]
\centering
\includegraphics[width=0.91\textwidth]{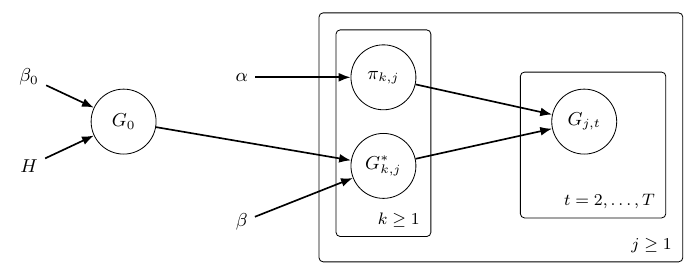}
\caption{Graphical representation of the model. The common base measure $G_0$ is drawn from a DP with concentration parameter $\beta_0$ and base measure $H$. For each departure state $j\in\mathcal{S}$, the atoms $G^*_{k,j}$, $k\geq1$, are drawn from a DP with concentration parameter $\beta$ and base measure $G_0$, while the weights $\pi_{k,j}$ follow a $GEM(\alpha)$ distribution. At each time $t=2,\dots,T$, the auxiliary variable $Z_{j,t}$ assigns the transition distribution $G_{j,t}$ to one of the atoms $G^*_{k,j}$.}
\label{fig:plate}
\end{figure}

Figure \ref{fig:plate} displays the dependency structure induced by the multidimensional HHDP through its graphical model representation. 

\subsection{The Generalized Infinite HMMs}\label{sec:exthmm}
In a longitudinal framework, let us consider a response variable $Y_{i,t} \in \mathbb{Y} \subseteq \mathbb{R}$, where $i=1,\ldots,N$, and $t=1,\ldots,T$. 
Given a suitable probability space we consider a doubly stochastic process $\{Y_{i,t}, V_{i,t}\}_{t=1}^T$ for $i=1,\ldots, N$. We assume that $\{Y_{i,t}\}_{t=1}^T$ is observable, while each $\{V_{i,t}\}_{t=1}^T$, with $i=1,\ldots,N$, is latent and follows a first-order Markov chain, where $V_{i,t} \in \mathbb{V} \subseteq \mathbb{N}$.

We denote by $\omega_{l}$ the initial probability $\mathbb{P}(V_{i,1}=l)$, while we denote by $\omega_{j, l}$ the transition probability $\mathbb{P}(V_{i,t} = l \lvert V_{i,t-1} = j)$, which are equal for all $i=1,\ldots,N$. In addition, we assume
\begin{equation*}\label{eq:1}
Y_{i,t} \lvert V_{i,t} = l \sim F(\theta_{l}), \quad l \in \mathbb{V},
\end{equation*}
where $F(\cdot)$ is the distribution of $Y_{i,t}$ given $V_{i,t} = l$ with parameter $\theta_{l}$. 
Notice that, based on the adopted distribution, we may have a parameter or a collection of parameters; for instance, if $F(\cdot)$ is a Gaussian distribution, we have $\mu_{l}$ and $\sigma^2_{l}$. 

A common assumption for HMMs is that of local independence, meaning that the variables $Y_{i,t}$ are conditionally independent given the latent process $\{V_{i,t}\}_{t=1}^T$. This implies that 
$$f(\boldsymbol{y} \lvert \boldsymbol{v}, \boldsymbol{\theta}) = \prod_{i=1}^N\prod_{t=1}^T f(y_{i,t} \lvert \theta_{v_{i,t}}),$$
where $\boldsymbol{y} = \{y_{1,1},\ldots y_{1,T}, \ldots ,y_{N,T}\}$ and $\boldsymbol{v} = \{v_{1,1},\ldots, v_{1,T}, \ldots v_{N,T}\}$ are the realizations of the stochastic processes.

 Following the approach proposed by \cite{Lijoi_2023} we can consider the MHHDP discussed in Section \ref{sec:ext}. Let us consider 
\begin{equation}\label{eq:HHDP_HMM1}
\{G_{j,2},\dots,G_{j,T}\}_{j\in\mathcal{S}} \sim MHHDP(\alpha,\beta,\beta_0;H),
\end{equation}
where
\begin{gather}
G_{j,t} \lvert Q_j \overset{\text{i.i.d.}}{\sim} Q_j, \quad Q_j = \sum_{k \geq 1} \pi_{k,j} \delta_{G_{k,j}^*}, \quad \{\pi_{k,j}\}_{k \geq 1} \sim GEM(\alpha), \notag \\ G_{k,j}^* \mid G_0 \overset{\text{i.i.d.}}{\sim} DP(\beta; G_0), \quad G_0 \sim DP(\beta_0;H), 
\end{gather}
with $\{\pi_{k,j}\}_{k \geq 1}$ and $\{G_{k,j}^*\}_{k \geq 1}$ independent. It is worth noting that the time index in Equation \eqref{eq:HHDP_HMM1} runs from $2$ to $T$, with $G_{j,t}$ denoting the transition kernel used by state $j$ to generate the transitions recorded at calendar time $t$; the motivation for this indexing choice, tied to the distinguished initial state, is explained in the following of this section. 
In Equation \eqref{GIHMM_def1}, $GEM(\alpha)$ governs how the (infinitely many) atoms $\{G_{k,j}^*\}_{k\geq1}$ are mixed into $Q_{j}$. In particular, a small $\alpha$ concentrates most of the mass on a few atoms, favoring ties among the transition kernels $G_{j,t}$ used at different time points $t$, that is, aggressive clustering of time points within state $j$, whereas a large $\alpha$ spreads the mass thinly over many atoms, making such ties increasingly unlikely. The same $GEM(\beta)$ and $GEM(\beta_0)$ notation will be used for the weights of $G_{k,j}^*$ and $G_0$, respectively, with an analogous interpretation at those lower levels of the hierarchy.
The infinite sequence $\{G_{k,j}^*\}_{k\geq1}$ is then represented as
\begin{gather*}
G_{k,j}^*=\sum_{l\geq 1} \omega_{k,j,l}\; \delta_{W_{k,j,l}} \quad W_{k,j,l}|G_0 \overset{i.i.d.}{\sim} G_0, \quad G_0 = \sum_{l \geq 1}\omega_{0,l}\; \delta_{\theta^*_l}, \quad \theta^*_l \overset{i.i.d.}{\sim} H, \\
\{\omega_{k,j,l}\}_{l \geq 1} \overset{i.i.d.}{\sim} GEM(\beta), \quad \{\omega_{0,l}\}_{l \geq 1} \sim GEM(\beta_0).
\end{gather*}
In this case independence is assumed between the sequences $\{\omega_{k,j,l}\}_{l \geq 1}$ and $\{W_{k,j,l}\}_{l \geq 1}$, and between $\{\omega_{0,l}\}_{l \geq 1}$ and $\{\theta^*_l\}_{l \geq 1}$. 
As defined in Section \ref{sec:ext}, where the atoms $\theta^*_l$ play the role of $X^*_l$, we may have
\begin{equation}\label{eq:G_star}
G_{k,j}^* = \sum_{l \geq 1} \omega^*_{k,j,l} \; \delta_{\theta^*_l} \quad \quad G_0 = \sum_{l \geq 1} \omega_{0,l} \; \delta_{\theta^*_l},
\end{equation}
where $\{\omega^*_{k,j,l}\}_{l \geq 1} \lvert \boldsymbol{\omega}_0 \overset{i.i.d.}{\sim} DP(\beta;\boldsymbol{\omega}_0)$, $\boldsymbol{\omega}_0 = \{\omega_{0,l}\}_{l \geq 1} \sim GEM(\beta_0)$, and $\theta^*_l\overset{i.i.d.}{\sim}H$ for $l \geq 1$. 

We refer to the HMM obtained from the previous formulation as GIHMM. In particular, the latent states and observational data can be modeled including an additional latent variable $Z_{j,t} \in \mathcal{Z} = \{1,2,\ldots\}$ that defines the cluster level for the $j$-th collection of transition probabilities, with $G_{j,t}=G^*_{k,j}$ whenever $Z_{j,t}=k$, $t=2,\ldots,T$, from which we have
$$Z_{j,t} \sim \boldsymbol{\pi}_j,$$
$$V_{i,t} \lvert V_{i,t-1} = j, Z_{j,t} = k \sim \boldsymbol{\omega}^*_{k,j},$$
$$Y_{i,t} \lvert V_{i,t} = l \sim F(\theta_{l}),$$
where $\boldsymbol{\pi}_j = \{\pi_{k,j}\}_{k \geq 1}$ is the collection of clustering probabilities defined in Equation \eqref{GIHMM_def1} for the $j$-th row $\boldsymbol{\omega}^*_{k,j} = \{\omega^*_{k,j,l}\}_{l \geq 1}$. In practice, $\pi_{k,j} = \mathbb{P}(Z_{j,t} = k)$, $t=2,\ldots,T$, while 
$$\omega^*_{k,j,l} = \mathbb{P}(V_{i,t} = l \lvert V_{i,t-1} = j, Z_{j,t} = k), \quad t=2,\ldots,T, \quad i=1,\ldots,N.$$ 

\begin{figure}[ht]
\centering
\resizebox{\textwidth}{!}{%
\begin{tikzpicture}[
  every node/.style={font=\small},
  cellval/.style={font=\scriptsize},
]
\definecolor{cA}{RGB}{27,158,119}
\definecolor{cB}{RGB}{217,95,2}
\definecolor{cC}{RGB}{117,112,179}
\definecolor{cD}{RGB}{231,41,138}
\definecolor{cE}{RGB}{102,102,102}
\definecolor{cF}{RGB}{166,86,40}
\node at (0.575,4.350) {\textbf{1}};
\node at (1.725,4.350) {\textbf{2}};
\node at (2.875,4.350) {\textbf{3}};
\node at (4.025,4.350) {\textbf{4}};
\node at (4.910,4.350) {$\cdots$};
\node at (-0.450,3.325) {\textbf{1}};
\draw[fill=cA!18,draw=cA!70,line width=0.5pt] (0.000,2.850) rectangle ++(1.150,0.950);
\node[cellval,text=cA!85!black] at (0.575,3.325) {0.50};
\draw[fill=cA!18,draw=cA!70,line width=0.5pt] (1.150,2.850) rectangle ++(1.150,0.950);
\node[cellval,text=cA!85!black] at (1.725,3.325) {0.28};
\draw[fill=cA!18,draw=cA!70,line width=0.5pt] (2.300,2.850) rectangle ++(1.150,0.950);
\node[cellval,text=cA!85!black] at (2.875,3.325) {0.15};
\draw[fill=cA!18,draw=cA!70,line width=0.5pt] (3.450,2.850) rectangle ++(1.150,0.950);
\node[cellval,text=cA!85!black] at (4.025,3.325) {0.05};
\draw[fill=cA!10,draw=cA!40,line width=0.4pt] (4.600,2.850) -- ++(0,0.950) -- ++(0.620,0) -- ++(0,-0.950);
\node[cellval,text=cA!60!black] at (4.910,3.325) {$\cdots$};
\node at (-0.450,2.375) {\textbf{2}};
\draw[fill=cB!18,draw=cB!70,line width=0.5pt] (0.000,1.900) rectangle ++(1.150,0.950);
\node[cellval,text=cB!85!black] at (0.575,2.375) {0.35};
\draw[fill=cB!18,draw=cB!70,line width=0.5pt] (1.150,1.900) rectangle ++(1.150,0.950);
\node[cellval,text=cB!85!black] at (1.725,2.375) {0.30};
\draw[fill=cB!18,draw=cB!70,line width=0.5pt] (2.300,1.900) rectangle ++(1.150,0.950);
\node[cellval,text=cB!85!black] at (2.875,2.375) {0.20};
\draw[fill=cB!18,draw=cB!70,line width=0.5pt] (3.450,1.900) rectangle ++(1.150,0.950);
\node[cellval,text=cB!85!black] at (4.025,2.375) {0.10};
\draw[fill=cB!10,draw=cB!40,line width=0.4pt] (4.600,1.900) -- ++(0,0.950) -- ++(0.620,0) -- ++(0,-0.950);
\node[cellval,text=cB!60!black] at (4.910,2.375) {$\cdots$};
\node at (-0.450,1.425) {\textbf{3}};
\draw[fill=cC!18,draw=cC!70,line width=0.5pt] (0.000,0.950) rectangle ++(1.150,0.950);
\node[cellval,text=cC!85!black] at (0.575,1.425) {0.60};
\draw[fill=cC!18,draw=cC!70,line width=0.5pt] (1.150,0.950) rectangle ++(1.150,0.950);
\node[cellval,text=cC!85!black] at (1.725,1.425) {0.20};
\draw[fill=cC!18,draw=cC!70,line width=0.5pt] (2.300,0.950) rectangle ++(1.150,0.950);
\node[cellval,text=cC!85!black] at (2.875,1.425) {0.12};
\draw[fill=cC!18,draw=cC!70,line width=0.5pt] (3.450,0.950) rectangle ++(1.150,0.950);
\node[cellval,text=cC!85!black] at (4.025,1.425) {0.05};
\draw[fill=cC!10,draw=cC!40,line width=0.4pt] (4.600,0.950) -- ++(0,0.950) -- ++(0.620,0) -- ++(0,-0.950);
\node[cellval,text=cC!60!black] at (4.910,1.425) {$\cdots$};
\node at (-0.450,0.475) {\textbf{4}};
\draw[fill=cD!18,draw=cD!70,line width=0.5pt] (0.000,0.000) rectangle ++(1.150,0.950);
\node[cellval,text=cD!85!black] at (0.575,0.475) {0.40};
\draw[fill=cD!18,draw=cD!70,line width=0.5pt] (1.150,0.000) rectangle ++(1.150,0.950);
\node[cellval,text=cD!85!black] at (1.725,0.475) {0.25};
\draw[fill=cD!18,draw=cD!70,line width=0.5pt] (2.300,0.000) rectangle ++(1.150,0.950);
\node[cellval,text=cD!85!black] at (2.875,0.475) {0.20};
\draw[fill=cD!18,draw=cD!70,line width=0.5pt] (3.450,0.000) rectangle ++(1.150,0.950);
\node[cellval,text=cD!85!black] at (4.025,0.475) {0.10};
\draw[fill=cD!10,draw=cD!40,line width=0.4pt] (4.600,0.000) -- ++(0,0.950) -- ++(0.620,0) -- ++(0,-0.950);
\node[cellval,text=cD!60!black] at (4.910,0.475) {$\cdots$};
\node at (-0.450,-0.275) {$\vdots$};
\draw[draw=gray!45,line width=0.4pt] (0.000,0.000) -- ++(0,-0.550) ++(0,0.550) -- ++(1.150,0);
\node[cellval,text=gray!70] at (0.575,-0.275) {$\vdots$};
\draw[draw=gray!45,line width=0.4pt] (1.150,0.000) -- ++(0,-0.550) ++(0,0.550) -- ++(1.150,0);
\node[cellval,text=gray!70] at (1.725,-0.275) {$\vdots$};
\draw[draw=gray!45,line width=0.4pt] (2.300,0.000) -- ++(0,-0.550) ++(0,0.550) -- ++(1.150,0);
\node[cellval,text=gray!70] at (2.875,-0.275) {$\vdots$};
\draw[draw=gray!45,line width=0.4pt] (3.450,0.000) -- ++(0,-0.550) ++(0,0.550) -- ++(1.150,0);
\node[cellval,text=gray!70] at (4.025,-0.275) {$\vdots$};
\node[cellval,text=gray!60] at (4.910,-0.275) {$\ddots$};
\node at (2.610,-1.100) {\textbf{$t=2$}};
\node at (8.395,4.350) {\textbf{1}};
\node at (9.545,4.350) {\textbf{2}};
\node at (10.695,4.350) {\textbf{3}};
\node at (11.845,4.350) {\textbf{4}};
\node at (12.730,4.350) {$\cdots$};
\node at (7.370,3.325) {\textbf{1}};
\draw[fill=cA!18,draw=cA!70,line width=0.5pt] (7.820,2.850) rectangle ++(1.150,0.950);
\node[cellval,text=cA!85!black] at (8.395,3.325) {0.50};
\draw[fill=cA!18,draw=cA!70,line width=0.5pt] (8.970,2.850) rectangle ++(1.150,0.950);
\node[cellval,text=cA!85!black] at (9.545,3.325) {0.28};
\draw[fill=cA!18,draw=cA!70,line width=0.5pt] (10.120,2.850) rectangle ++(1.150,0.950);
\node[cellval,text=cA!85!black] at (10.695,3.325) {0.15};
\draw[fill=cA!18,draw=cA!70,line width=0.5pt] (11.270,2.850) rectangle ++(1.150,0.950);
\node[cellval,text=cA!85!black] at (11.845,3.325) {0.05};
\draw[fill=cA!10,draw=cA!40,line width=0.4pt] (12.420,2.850) -- ++(0,0.950) -- ++(0.620,0) -- ++(0,-0.950);
\node[cellval,text=cA!60!black] at (12.730,3.325) {$\cdots$};
\node at (7.370,2.375) {\textbf{2}};
\draw[fill=cE!18,draw=cE!70,line width=0.5pt] (7.820,1.900) rectangle ++(1.150,0.950);
\node[cellval,text=cE!85!black] at (8.395,2.375) {0.45};
\draw[fill=cE!18,draw=cE!70,line width=0.5pt] (8.970,1.900) rectangle ++(1.150,0.950);
\node[cellval,text=cE!85!black] at (9.545,2.375) {0.30};
\draw[fill=cE!18,draw=cE!70,line width=0.5pt] (10.120,1.900) rectangle ++(1.150,0.950);
\node[cellval,text=cE!85!black] at (10.695,2.375) {0.15};
\draw[fill=cE!18,draw=cE!70,line width=0.5pt] (11.270,1.900) rectangle ++(1.150,0.950);
\node[cellval,text=cE!85!black] at (11.845,2.375) {0.07};
\draw[fill=cE!10,draw=cE!40,line width=0.4pt] (12.420,1.900) -- ++(0,0.950) -- ++(0.620,0) -- ++(0,-0.950);
\node[cellval,text=cE!60!black] at (12.730,2.375) {$\cdots$};
\node at (7.370,1.425) {\textbf{3}};
\draw[fill=cC!18,draw=cC!70,line width=0.5pt] (7.820,0.950) rectangle ++(1.150,0.950);
\node[cellval,text=cC!85!black] at (8.395,1.425) {0.60};
\draw[fill=cC!18,draw=cC!70,line width=0.5pt] (8.970,0.950) rectangle ++(1.150,0.950);
\node[cellval,text=cC!85!black] at (9.545,1.425) {0.20};
\draw[fill=cC!18,draw=cC!70,line width=0.5pt] (10.120,0.950) rectangle ++(1.150,0.950);
\node[cellval,text=cC!85!black] at (10.695,1.425) {0.12};
\draw[fill=cC!18,draw=cC!70,line width=0.5pt] (11.270,0.950) rectangle ++(1.150,0.950);
\node[cellval,text=cC!85!black] at (11.845,1.425) {0.05};
\draw[fill=cC!10,draw=cC!40,line width=0.4pt] (12.420,0.950) -- ++(0,0.950) -- ++(0.620,0) -- ++(0,-0.950);
\node[cellval,text=cC!60!black] at (12.730,1.425) {$\cdots$};
\node at (7.370,0.475) {\textbf{4}};
\draw[fill=cD!18,draw=cD!70,line width=0.5pt] (7.820,0.000) rectangle ++(1.150,0.950);
\node[cellval,text=cD!85!black] at (8.395,0.475) {0.40};
\draw[fill=cD!18,draw=cD!70,line width=0.5pt] (8.970,0.000) rectangle ++(1.150,0.950);
\node[cellval,text=cD!85!black] at (9.545,0.475) {0.25};
\draw[fill=cD!18,draw=cD!70,line width=0.5pt] (10.120,0.000) rectangle ++(1.150,0.950);
\node[cellval,text=cD!85!black] at (10.695,0.475) {0.20};
\draw[fill=cD!18,draw=cD!70,line width=0.5pt] (11.270,0.000) rectangle ++(1.150,0.950);
\node[cellval,text=cD!85!black] at (11.845,0.475) {0.10};
\draw[fill=cD!10,draw=cD!40,line width=0.4pt] (12.420,0.000) -- ++(0,0.950) -- ++(0.620,0) -- ++(0,-0.950);
\node[cellval,text=cD!60!black] at (12.730,0.475) {$\cdots$};
\node at (7.370,-0.275) {$\vdots$};
\draw[draw=gray!45,line width=0.4pt] (7.820,0.000) -- ++(0,-0.550) ++(0,0.550) -- ++(1.150,0);
\node[cellval,text=gray!70] at (8.395,-0.275) {$\vdots$};
\draw[draw=gray!45,line width=0.4pt] (8.970,0.000) -- ++(0,-0.550) ++(0,0.550) -- ++(1.150,0);
\node[cellval,text=gray!70] at (9.545,-0.275) {$\vdots$};
\draw[draw=gray!45,line width=0.4pt] (10.120,0.000) -- ++(0,-0.550) ++(0,0.550) -- ++(1.150,0);
\node[cellval,text=gray!70] at (10.695,-0.275) {$\vdots$};
\draw[draw=gray!45,line width=0.4pt] (11.270,0.000) -- ++(0,-0.550) ++(0,0.550) -- ++(1.150,0);
\node[cellval,text=gray!70] at (11.845,-0.275) {$\vdots$};
\node[cellval,text=gray!60] at (12.730,-0.275) {$\ddots$};
\node at (10.430,-1.100) {\textbf{$t=3$}};
\node at (16.215,4.350) {\textbf{1}};
\node at (17.365,4.350) {\textbf{2}};
\node at (18.515,4.350) {\textbf{3}};
\node at (19.665,4.350) {\textbf{4}};
\node at (20.550,4.350) {$\cdots$};
\node at (15.190,3.325) {\textbf{1}};
\draw[fill=cA!18,draw=cA!70,line width=0.5pt] (15.640,2.850) rectangle ++(1.150,0.950);
\node[cellval,text=cA!85!black] at (16.215,3.325) {0.50};
\draw[fill=cA!18,draw=cA!70,line width=0.5pt] (16.790,2.850) rectangle ++(1.150,0.950);
\node[cellval,text=cA!85!black] at (17.365,3.325) {0.28};
\draw[fill=cA!18,draw=cA!70,line width=0.5pt] (17.940,2.850) rectangle ++(1.150,0.950);
\node[cellval,text=cA!85!black] at (18.515,3.325) {0.15};
\draw[fill=cA!18,draw=cA!70,line width=0.5pt] (19.090,2.850) rectangle ++(1.150,0.950);
\node[cellval,text=cA!85!black] at (19.665,3.325) {0.05};
\draw[fill=cA!10,draw=cA!40,line width=0.4pt] (20.240,2.850) -- ++(0,0.950) -- ++(0.620,0) -- ++(0,-0.950);
\node[cellval,text=cA!60!black] at (20.550,3.325) {$\cdots$};
\node at (15.190,2.375) {\textbf{2}};
\draw[fill=cE!18,draw=cE!70,line width=0.5pt] (15.640,1.900) rectangle ++(1.150,0.950);
\node[cellval,text=cE!85!black] at (16.215,2.375) {0.45};
\draw[fill=cE!18,draw=cE!70,line width=0.5pt] (16.790,1.900) rectangle ++(1.150,0.950);
\node[cellval,text=cE!85!black] at (17.365,2.375) {0.30};
\draw[fill=cE!18,draw=cE!70,line width=0.5pt] (17.940,1.900) rectangle ++(1.150,0.950);
\node[cellval,text=cE!85!black] at (18.515,2.375) {0.15};
\draw[fill=cE!18,draw=cE!70,line width=0.5pt] (19.090,1.900) rectangle ++(1.150,0.950);
\node[cellval,text=cE!85!black] at (19.665,2.375) {0.07};
\draw[fill=cE!10,draw=cE!40,line width=0.4pt] (20.240,1.900) -- ++(0,0.950) -- ++(0.620,0) -- ++(0,-0.950);
\node[cellval,text=cE!60!black] at (20.550,2.375) {$\cdots$};
\node at (15.190,1.425) {\textbf{3}};
\draw[fill=cC!18,draw=cC!70,line width=0.5pt] (15.640,0.950) rectangle ++(1.150,0.950);
\node[cellval,text=cC!85!black] at (16.215,1.425) {0.60};
\draw[fill=cC!18,draw=cC!70,line width=0.5pt] (16.790,0.950) rectangle ++(1.150,0.950);
\node[cellval,text=cC!85!black] at (17.365,1.425) {0.20};
\draw[fill=cC!18,draw=cC!70,line width=0.5pt] (17.940,0.950) rectangle ++(1.150,0.950);
\node[cellval,text=cC!85!black] at (18.515,1.425) {0.12};
\draw[fill=cC!18,draw=cC!70,line width=0.5pt] (19.090,0.950) rectangle ++(1.150,0.950);
\node[cellval,text=cC!85!black] at (19.665,1.425) {0.05};
\draw[fill=cC!10,draw=cC!40,line width=0.4pt] (20.240,0.950) -- ++(0,0.950) -- ++(0.620,0) -- ++(0,-0.950);
\node[cellval,text=cC!60!black] at (20.550,1.425) {$\cdots$};
\node at (15.190,0.475) {\textbf{4}};
\draw[fill=cF!18,draw=cF!70,line width=0.5pt] (15.640,0.000) rectangle ++(1.150,0.950);
\node[cellval,text=cF!85!black] at (16.215,0.475) {0.55};
\draw[fill=cF!18,draw=cF!70,line width=0.5pt] (16.790,0.000) rectangle ++(1.150,0.950);
\node[cellval,text=cF!85!black] at (17.365,0.475) {0.22};
\draw[fill=cF!18,draw=cF!70,line width=0.5pt] (17.940,0.000) rectangle ++(1.150,0.950);
\node[cellval,text=cF!85!black] at (18.515,0.475) {0.14};
\draw[fill=cF!18,draw=cF!70,line width=0.5pt] (19.090,0.000) rectangle ++(1.150,0.950);
\node[cellval,text=cF!85!black] at (19.665,0.475) {0.06};
\draw[fill=cF!10,draw=cF!40,line width=0.4pt] (20.240,0.000) -- ++(0,0.950) -- ++(0.620,0) -- ++(0,-0.950);
\node[cellval,text=cF!60!black] at (20.550,0.475) {$\cdots$};
\node at (15.190,-0.275) {$\vdots$};
\draw[draw=gray!45,line width=0.4pt] (15.640,0.000) -- ++(0,-0.550) ++(0,0.550) -- ++(1.150,0);
\node[cellval,text=gray!70] at (16.215,-0.275) {$\vdots$};
\draw[draw=gray!45,line width=0.4pt] (16.790,0.000) -- ++(0,-0.550) ++(0,0.550) -- ++(1.150,0);
\node[cellval,text=gray!70] at (17.365,-0.275) {$\vdots$};
\draw[draw=gray!45,line width=0.4pt] (17.940,0.000) -- ++(0,-0.550) ++(0,0.550) -- ++(1.150,0);
\node[cellval,text=gray!70] at (18.515,-0.275) {$\vdots$};
\draw[draw=gray!45,line width=0.4pt] (19.090,0.000) -- ++(0,-0.550) ++(0,0.550) -- ++(1.150,0);
\node[cellval,text=gray!70] at (19.665,-0.275) {$\vdots$};
\node[cellval,text=gray!60] at (20.550,-0.275) {$\ddots$};
\node at (18.250,-1.100) {\textbf{$t=4$}};
\draw[cA!65,dashed,line width=0.7pt] (5.280,3.325) to node[midway,above,font=\tiny,text=cA!70!black]{$=$} (7.200,3.325);
\draw[cC!65,dashed,line width=0.7pt] (5.280,1.425) to node[midway,above,font=\tiny,text=cC!70!black]{$=$} (7.200,1.425);
\draw[cD!65,dashed,line width=0.7pt] (5.280,0.475) to node[midway,above,font=\tiny,text=cD!70!black]{$=$} (7.200,0.475);
\draw[cA!65,dashed,line width=0.7pt] (13.100,3.325) to node[midway,above,font=\tiny,text=cA!70!black]{$=$} (15.020,3.325);
\draw[cE!65,dashed,line width=0.7pt] (13.100,2.375) to node[midway,above,font=\tiny,text=cE!70!black]{$=$} (15.020,2.375);
\draw[cC!65,dashed,line width=0.7pt] (13.100,1.425) to node[midway,above,font=\tiny,text=cC!70!black]{$=$} (15.020,1.425);
\end{tikzpicture}
}
\caption{Illustration of the clustering induced across three consecutive transition matrices. Each row displays the first four entries of the simplex governing the transitions out of departure state $j$ recorded at time $t$; within each matrix, all rows are distinct, while a row is allowed to coincide with another (same color and values). Dashed connectors highlight such equalities.}
\label{fig:transition_matrices}
\end{figure}
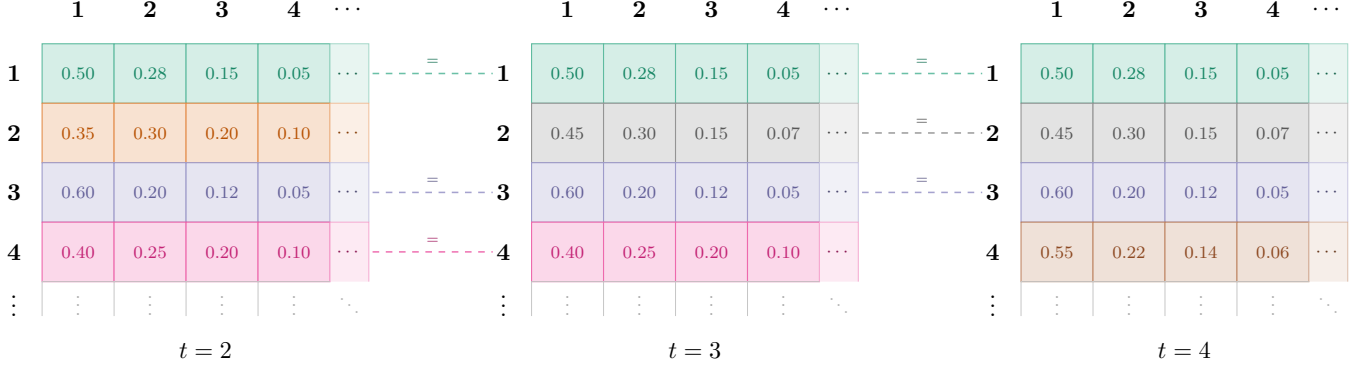

The proposed GIHMM provides a flexible framework that lies between the homogeneous and heterogeneous HMM formulations. In addition, the model allows the rows of the transition probability matrix to be clustered. Specifically, at each time point $t$, it defines a transition probability matrix whose rows may change over time. Notice that, if the realization of $Z_{j,t}$ denoted as $z_{j,t}$ is equal to 1 for all $j,t$, then we have the classical form of the infinite HMM. Figure \ref{fig:transition_matrices} illustrates the resulting clustering of the transition matrices over time.

The model introduced may be estimated considering the finite approximation of the DP \citep{muliere1998approximating,ishwaran2001gibbs}. However, 
instead of truncating the stick-breaking representation of the DP, we use the finite Dirichlet approximation proposed by \cite{ishwaran2002exact}. In particular, we set $j=1,\ldots,J$, $k=1,\ldots,K$, and $l=1,\ldots,L$. In addition, we constrain $J=L$. Let $\boldsymbol{\theta}^* = \{\theta^*_l\}_{l=1}^L$, $\boldsymbol{\pi} = \{\boldsymbol{\pi}_j\}_{j=1}^J$, $\boldsymbol{\omega}^* = \{\boldsymbol{\omega}^*_{k,j}\}_{k,j=1}^{K,J}$, $\boldsymbol{\omega}_0 = \{\omega_{0,l}\}_{l=1}^L$, $\boldsymbol{Z} = \{Z_{j,t}\}_{j=1, t=2}^{J,T}$, $\boldsymbol{V} = \{V_{i,t}\}_{i,t=1}^{N,T}$, and $\boldsymbol{Y} = \{Y_{i,t}\}_{i,t=1}^{N,T}$. We assume for simplicity that there is a distinguished initial state as in \cite{Teh_2006}, and this implies that the $\{G_{j,t}\}$ in Equation \eqref{eq:HHDP_HMM1} are indexed by $t=2,\ldots,T$. For simplicity the initial probability is assumed to be known.
Then, we can define the following augmented posterior distribution
\begin{equation}\label{eq:post}
p(\boldsymbol{\theta}^*, \boldsymbol{\pi}, \boldsymbol{\omega}^*, \boldsymbol{\omega}_0, \boldsymbol{z}, \boldsymbol{v} \lvert \boldsymbol{y}) \propto p(\boldsymbol{y} \lvert \boldsymbol{\theta}^*, \boldsymbol{v}) p(\boldsymbol{v} \lvert \boldsymbol{z}, \boldsymbol{\omega}^*) p(\boldsymbol{z} \lvert \boldsymbol{\pi}) p(\boldsymbol{\omega}^* \lvert \boldsymbol{\omega}_0) p(\boldsymbol{\pi}) 
p(\boldsymbol{\omega}_0) p(\boldsymbol{\theta}^*),
\end{equation}
where $\boldsymbol{z}$, $\boldsymbol{v}$, and $\boldsymbol{y}$ are the realizations of $\boldsymbol{Z}$, $\boldsymbol{V}$, and $\boldsymbol{Y}$, respectively. We assume 
$$\boldsymbol{\pi}_j \sim \mathcal{D}ir(\boldsymbol{\alpha}/K), \quad j =1, \ldots,J,$$
where $\boldsymbol{\alpha}$ is a vector of dimension $K$ containing the same $\alpha$. Then, we assume
$$\boldsymbol{\omega}^*_{k,j} \lvert \boldsymbol{\omega}_0 \sim \mathcal{D}ir(\beta \boldsymbol{\omega}_0), \quad k=1,\ldots,K, \quad j=1, \ldots,J,$$
and
$$\boldsymbol{\omega}_0 \sim \mathcal{D}ir(\boldsymbol{\beta}_0/L),$$
where $\boldsymbol{\beta}_0$ is a vector of dimension $L$ containing the same $\beta_0$. Regarding $\boldsymbol{\theta}^*$, its specification depends on the parameters of the distribution $F(\cdot)$. Accordingly, for $l = 1,\ldots,L$, we may assume
$\theta^*_l \sim \mathcal{L},$
where $\mathcal{L}$ denotes an appropriate prior distribution. 

In the following, we consider a multivariate Gaussian GIHMM, which means, $\boldsymbol{Y}_{i,t} \in \mathbb{R}^d$ and $F(\cdot) =\mathcal{N}(\boldsymbol{\mu}^*_l, \boldsymbol{\Sigma}^*_l)$. In this case, we assume 
$$\boldsymbol{\mu}^*_l \sim \mathcal{N}(\boldsymbol{m}, \boldsymbol{M}), \quad l = 1,\ldots,L,$$
while
$$\boldsymbol{\Sigma}^*_l \sim \mathcal{IW}(s, \boldsymbol{S}), \quad l = 1,\ldots,L,$$
where $\mathcal{IW}(\cdot, \cdot)$ denotes an Inverse-Wishart distribution.

\section{MCMC algorithm}\label{sec:4}

In this section, we describe the MCMC algorithm implemented for the estimation of the proposed model. In particular, we consider a Metropolis-within-Gibbs algorithm.

\subsection{Metropolis-within-Gibbs MCMC}

The MCMC here implemented follows the algorithm proposed by \cite{Lijoi_2023}. We use the augmented posterior distribution \eqref{eq:post},so that it is possible to derive the following full conditional distributions; see Appendix A for the proofs. 
In particular, for $\boldsymbol{\pi}_j$ we have
\begin{equation}\label{eq:ful_pi}
\boldsymbol{\pi}_j \lvert \cdots \sim \mathcal{D}ir(\boldsymbol{\alpha}/K + \boldsymbol{m}_j), \quad j=1,\ldots,J,
\end{equation}
where $\boldsymbol{m}_j = \left[ \sum_{t=2}^T \mathds{1}(z_{j,t} =1), \ldots, \sum_{t=2}^T \mathds{1}(z_{j,t} =K) \right]'$ with $\mathds{1}(\cdot)$ is an indicator function. For $\boldsymbol{\omega}_{k,j}$ we have
\begin{equation}\label{eq:ful_omega}
\boldsymbol{\omega}^*_{k,j} \lvert \cdots \sim \mathcal{D}ir(\beta\boldsymbol{\omega}_0 + \boldsymbol{n}_{k,j}), \quad k=1,\ldots,K, \quad j=1,\ldots,J,
\end{equation}
where 
$$\boldsymbol{n}_{k,j} = \left[ \sum_{t=2}^T \mathds{1}(z_{j,t} =k) \sum_{i=1}^N \mathds{1}(v_{i,t-1} = j, v_{i,t} = 1), \ldots, \sum_{t=2}^T \mathds{1}(z_{j,t} =k) \sum_{i=1}^N \mathds{1}(v_{i,t-1} = j, v_{i,t} = L) \right]'.$$  
Additionally, for $V_{i,t}$ we have
\begin{equation}\label{eq:ful_v}
p(V_{i,t} = l \lvert \cdots) \propto \omega^*_{z_{v_{i,t-1},t}; v_{i,t-1}; l} \times p(y_{i,t} \lvert \boldsymbol{\mu}^*_l, \boldsymbol{\Sigma}^*_l) \times \omega^*_{z_{l,t+1};l;v_{i,t+1}}, \quad l=1,\ldots,L,
\end{equation}
for $1<t<T$, where $\omega^*_{z_{v_{i,t-1},t}; v_{i,t-1}; l}$ is $\omega^*_{k,j,l}$ when $k=z_{v_{i,t-1},t}$, and $j=v_{i,t-1}$, while $\omega^*_{z_{l,t+1};l;v_{i,t+1}}$ is $\omega^*_{k,j,l}$ when $k=z_{l,t+1}$, $j=l$, and $l=v_{i,t+1}$. Notice that, similar forms are obtained for $t=1$ and $t=T$. We use $``;"$ instead of $``,"$ to improve readability. 

Regarding $Z_{j,t}$, we have
\begin{equation}\label{eq:ful_z}
p(Z_{j,t} = k \lvert \cdots) \propto \pi_{k,j} \prod_{l=1}^L \prod_{i=1}^N \omega_{k,j,l}^{* \mathds{1}(v_{i,t-1} = j, v_{i,t} = l)}, 
\end{equation}
for $k=1,\ldots,K$, $j=1,\ldots,J$, and $t=2,\ldots,T$. For the mean vectors we obtain
\begin{equation}
\boldsymbol{\mu}^*_l \lvert \cdots  \sim \mathcal{N} \left( \tilde{\boldsymbol{M}}_l
 \tilde{\boldsymbol{m}}_l,\tilde{\boldsymbol{M}}_l \right), \quad l=1,\ldots,L,
  \label{eq:11}
\end{equation}  
where
$$\tilde{\boldsymbol{M}}_l^{-1} = 
n_l \boldsymbol{\Sigma}_l^{*-1}+ \boldsymbol{M}^{-1} \quad \text{and} \quad
\tilde{\boldsymbol{m}}_l = 
\boldsymbol{\Sigma}_l^{*-1} \boldsymbol{\tilde{y}}_l 
+ \boldsymbol{M}^{-1}\boldsymbol{m}
, $$
with $n_l = \sum_{i=1}^N \sum_{t=1}^{T} \mathds{1}(V_{i,t} = l)$ and ${\boldsymbol{\tilde{y}}_l} =  
\sum_{i=1}^N \sum_{t=1}^{T} \boldsymbol{y}_{i,t} \mathds{1}(V_{i,t} = l).$

Finally, for the covariance matrices we have 
\begin{equation}
\boldsymbol{\Sigma}^*_l \lvert \cdots  \sim \mathcal{IW}(s + n_l, \boldsymbol{S} + \tilde{\boldsymbol{S}}_l), \quad l=1,\ldots,L,
\label{eq:12} 
\end{equation}
where
$$\tilde{\boldsymbol{S}}_l = \sum_{i=1}^N \sum_{t=1}^{T} (\boldsymbol{y}_{i,t} -\boldsymbol{\mu}^*_l)(\boldsymbol{y}_{i,t}-\boldsymbol{\mu}^*_l)' \mathds{1}(V_{i,t} = l).$$

For $\boldsymbol{\omega}_0$ we use a Metropolis-Hastings (MH) algorithm \citep{metropolis1953equation,hastings1970monte}, in particular we consider a random-walk MH algorithm using the same transformation adopted in \cite{Lijoi_2023}, that is, $\log{(\omega_{0,1}/\omega_{0,L})}, \ldots, \log{(\omega_{0,L-1}/\omega_{0,L})}$.

\subsection{Comments on the complexity of the model}

The GIHMM proposed in this work is a flexible but computationally challenging model, as it involves two distinct levels of latent structures. The first level concerns the clustering of the rows of the transition probability matrix across time, while the second corresponds to the latent states of the HMM, which describe the evolution of the underlying process over time. From a computational perspective, a possible alternative strategy would be to consider a fixed formulation, where both the number of latent states and the number of clusters associated with each row of the transition matrix are specified \textit{a priori}. However, such an approach would require fitting multiple models corresponding to all possible combinations of the number of states and the number of transition clusters. In particular, for a certain number of latent states $(J)$ and a certain number of clusters for the transition rows $(K)$, this strategy would require $J \times K$ separate MCMC algorithms, followed by a model comparison step to select the most appropriate specification, apart from the case in which a reduced number of combination is considered on the basis of a suitable grid. Obviously, this procedure would be computationally demanding and would not account for the uncertainty associated with the unknown number of states and clusters.

Differently from using many separate MCMC algorithms, the nonparametric formulation adopted in the GIHMM allows Bayesian inference by a single MCMC algorithm. The hierarchical structure of the model enables the number of latent states and the clustering of transition rows to be directly obtained from the data, avoiding the need to repeatedly fit different models. 

 \section{Synthetic and real data applications}\label{sec:5}
 
In this section, we initially consider simulated data. In particular, we investigate two scenarios in which the data are generated from either a homogeneous or a non-homogeneous HMM.  Finally, we present two real-data applications. In particular, the first dataset consists of a multivariate analysis of world development indicators, while the second we analyze the state-level production data. The {\tt R} codes used for the simulation study and the real-data applications are provided in the supplementary material.

\subsection{Synthetic dataset 1: The Heterogeneous Case}\label{sec:sim1}

We consider a general setting with $N=30$ units, $T=30$ time points, and a true number of latent states equal to 2. In this simulated dataset, observations are generated from a model with time-varying transition dynamics. Specifically, the transition probability matrix is assumed to be constant during the first 15 time points and is then replaced by a different transition probability matrix for the remaining time points. This setting defines a structural change in the latent dynamics, allowing us to assess the ability of the proposed GIHMM to detect and adapt to temporal heterogeneity in the transition mechanism. This simulation provides a challenging benchmark for evaluating whether the GIHMM can accurately recover both the latent state sequence and the evolving transition structure, while we expect that the infinite HMM, when applied to the heterogeneous case, may tend to overestimate the number of components.

Regarding the assumed parameter values, the first transition matrix represents a situation in which transitions between states are more likely, whereas the second matrix corresponds to a setting where the process is more persistent, with a higher probability of remaining in the same state. Specifically, the transition matrices for the first and last time points are, respectively,
\[
\begin{bmatrix}
0.31 & 0.69\\
0.79 & 0.21
\end{bmatrix}
\quad \text{and} \quad
\begin{bmatrix}
0.69 & 0.31\\
0.21 & 0.79
\end{bmatrix}.
\]
The transition matrices reported above are obtained by considering two clusters for the latent variables $Z_{j,t}$, $j=1,2$. Specifically, for $j=1$, the first row of the transition matrix on the left corresponds to the first cluster, whereas the first row of the matrix on the right corresponds to the second cluster. The same interpretation applies to $j=2$ considering the second rows of the two matrices. 

Regarding the vector means, we consider $[35,35]'$ and $[40,40]'$ while for the variance-covariance matrices we use an identity matrix $\boldsymbol{I}$.
We set $J=L=10$ and $K=5$, while for the hyperparameters we impose $\beta = 1, \alpha = 1, \beta_0 = 1, s = 6, \boldsymbol{S} = s \times \boldsymbol{I}, \boldsymbol{m} = \boldsymbol{0}, \boldsymbol{M} = 100 \times \boldsymbol{I}$, where $\boldsymbol{I}$ is the identity matrix. Both algorithms are run for 20,000 MCMC iterations, with the first $50\%$ discarded as burn-in. No thinning is applied.

\begin{figure}[!htbp]
\centering
\includegraphics[width=0.7\textwidth]{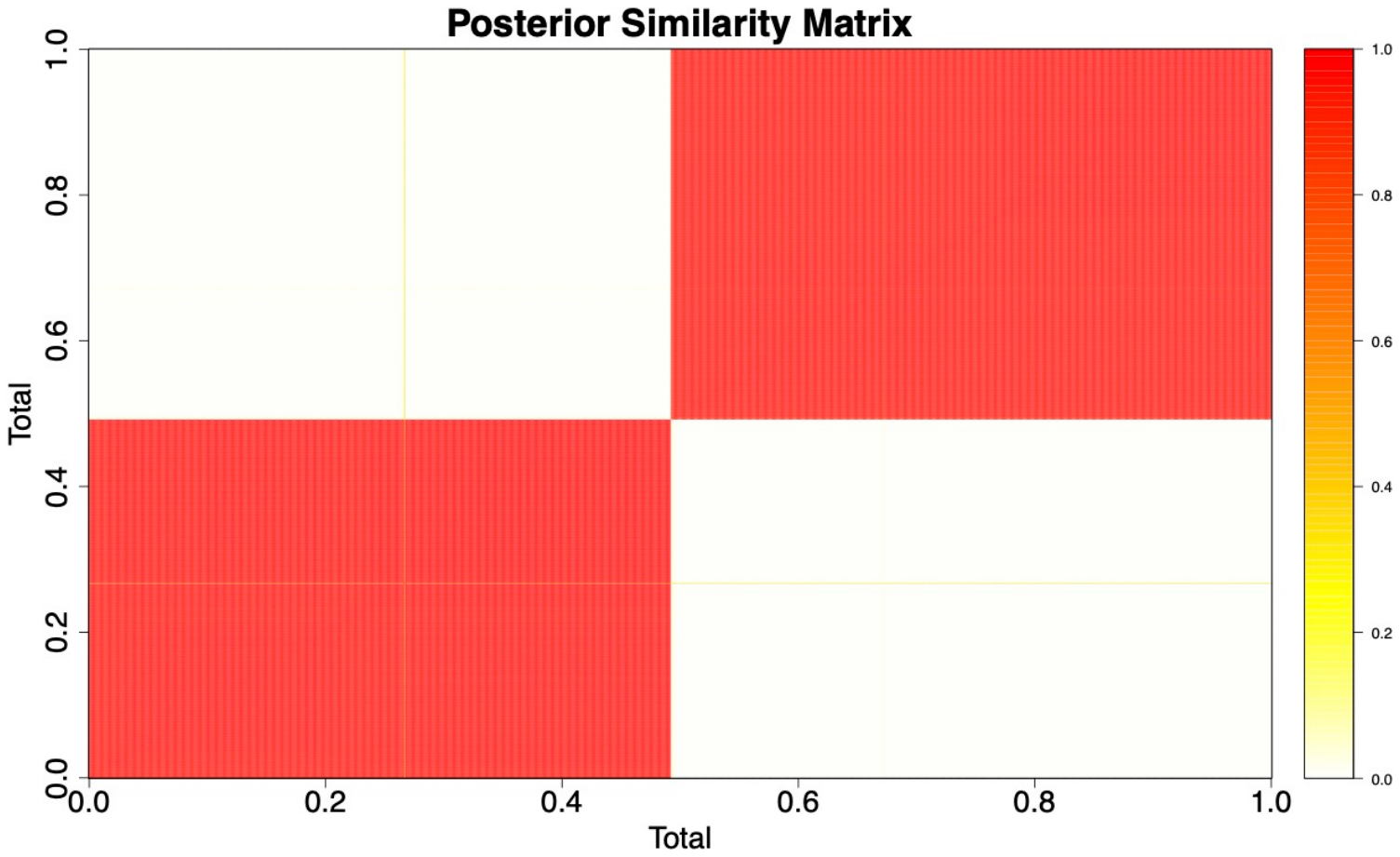}
\caption{Synthetic dataset 1. Posterior similarity matrix for $\boldsymbol{V}$ of the GIHMM. Total, on the $x$- and $y$- axes, means that all the $V_{i,t}$ are considered.}
\label{fig1}
\end{figure}

We fit the GIHMM and the infinite HMM to the synthetic dataset and we compare their performance. For each model, the final partition of the latent states is obtained by minimizing the Variation of Information (VI) loss \citep{meilua2007comparing}. The GIHMM identifies two latent states, which is the true number of states, whereas the infinite HMM identifies three latent states. To further assess the latent structure, we also examine the posterior similarity matrix for each model. In particular, the posterior similarity matrix for the GIHMM is shown in Figure \ref{fig1}.

Figure \ref{fig1} clearly shows the presence of two well-separated clusters, indicating that the GIHMM successfully recovers the underlying latent-state structure. 

\begin{figure}[!htbp]
\centering
\includegraphics[width=0.7\textwidth]{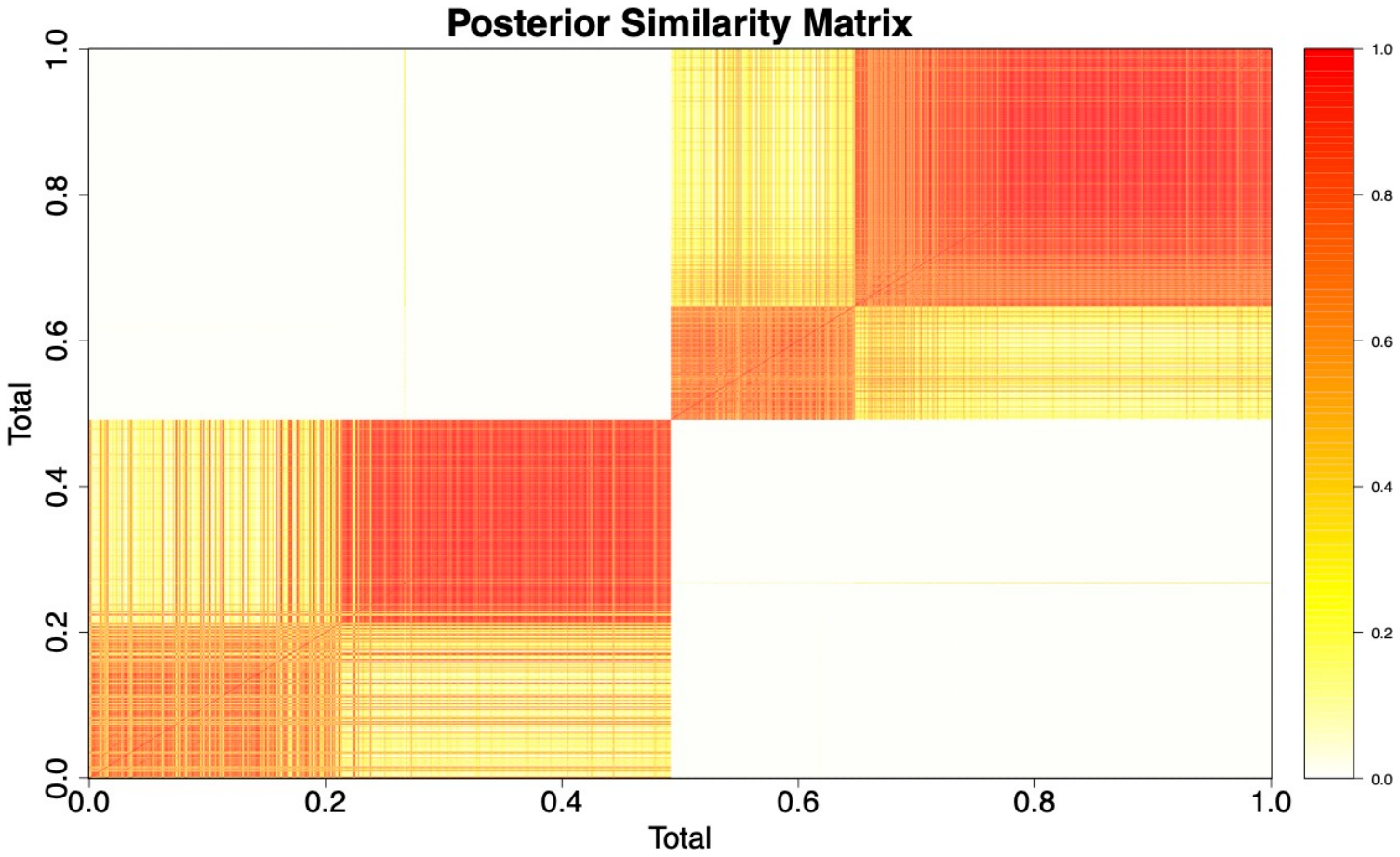}
\caption{Synthetic dataset 1. Posterior similarity matrix for $\boldsymbol{V}$ of the infinite HMM. Total, on the $x$- and $y$- axes, means that all the $V_{i,t}$ are considered.}
\label{fig2}
\end{figure}

The posterior similarity matrix for the infinite HMM is represented in Figure \ref{fig2}.

Differently from Figure \ref{fig1}, Figure \ref{fig2} shows evidence of more than two distinct states, indicating that the infinite HMM overestimates the number of latent states in this setting. This suggests that, when the transition probability matrix varies over time, the infinite HMM may accommodate the resulting heterogeneity by introducing additional states rather than capturing the variation in the transition dynamics. 

We now provide a more detailed analysis of the results obtained for the GIHMM. The model correctly identifies the change of the transition matrix, using the clusters of the rows of the transition matrix, identifying two clusters for each row. In addition, we evaluate the classification accuracy of the model by comparing the estimated latent states with the true underlying states. The classification accuracy is computed as the proportion of correctly classified latent states over the total number of latent variables. Table \ref{tab:1} reports the classification accuracy for the two latent structures involved in the GIHMM, which are $\boldsymbol{V}$ and $\boldsymbol{Z}$.

\begin{table}[ht]
\centering
\caption{Classification accuracy for the GIHMM.}
\label{tab:1}
\begin{tabular}{lr}
\hline
Latent structure & Accuracy $(\%)$ \\
\hline
$\boldsymbol{V}$ & 99.89 \\
$\boldsymbol{Z}$ & 96.55 \\
\hline
\end{tabular}
\end{table}

\begin{figure}[!htbp]
\centering
\includegraphics[width=0.7\textwidth]{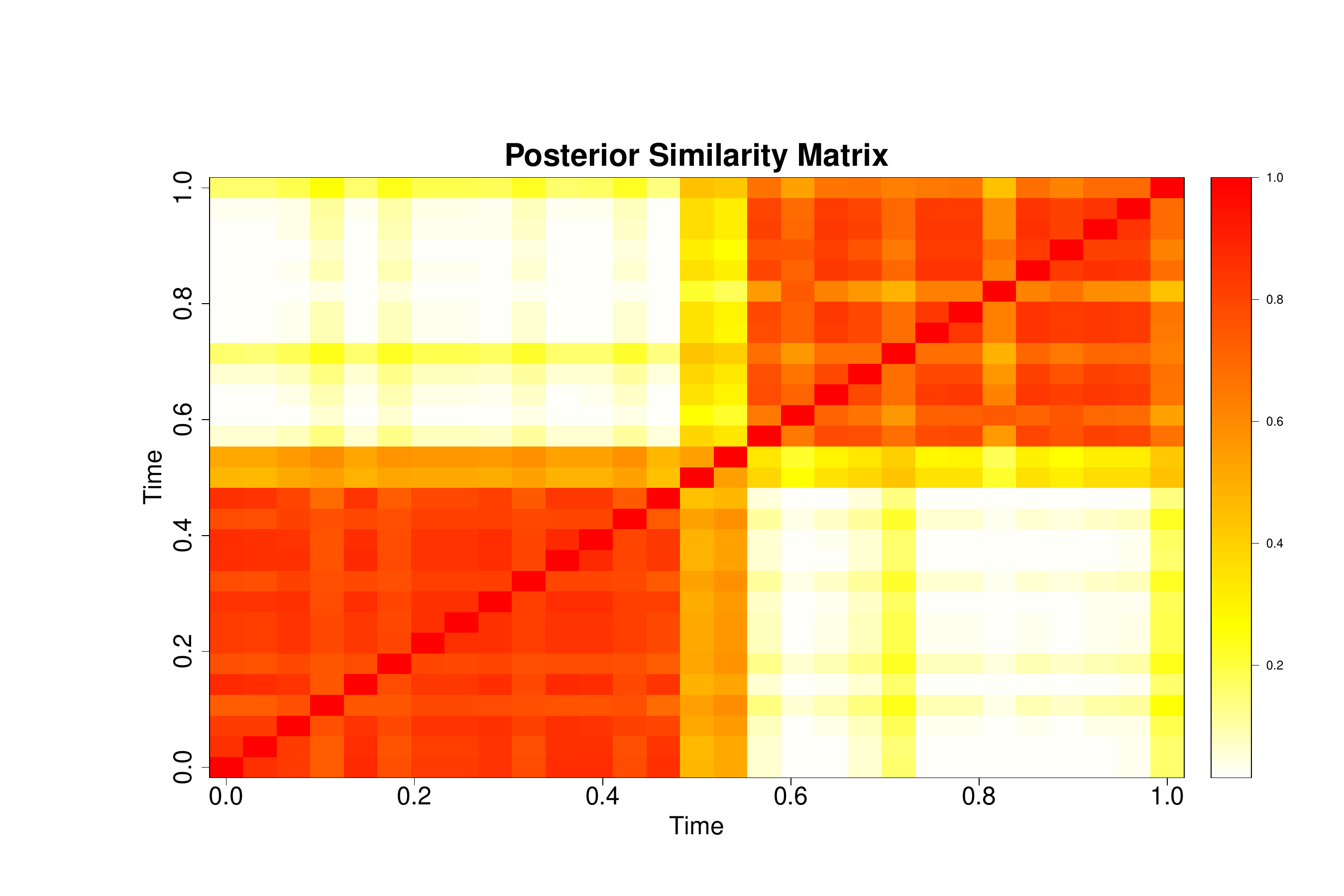}
\caption{Synthetic dataset 1. Posterior similarity matrix for $\boldsymbol{Z}$, first row, of the GIHMM.}
\label{fig3}
\end{figure}

Looking at Table \ref{tab:1}, we observe that the model achieves an accuracy of $99.89\%$ for the latent states $\boldsymbol{V}$, indicating an almost perfect recovery of the true hidden-state sequence. Similarly, the latent structure $\boldsymbol{Z}$ is accurately identified, with a classification accuracy of $96.55\%$. These results confirm the ability of the GIHMM to correctly recover both the evolution of the latent states and the clustering structure of the transition probability matrices. Similarly, for the variables in $\boldsymbol{Z}$, we can analyze the clustering structure by considering the posterior similarity matrices associated with different values of $j$. In particular, by partitioning the latent variables $Z_{j,t}$ according to $j=1,2$, we obtain the corresponding posterior similarity matrices, which is reported in Figure \ref{fig3}.

Regarding the model parameters, we divide the discussion into two parts. In particular, Table \ref{tab:2} reports the posterior estimates for the gaussian emission parameters, whereas Table \ref{tab:3} presents the results for the parameters governing the latent processes. The posterior means of these parameters are computed using MCMC draws selected according to the final configuration of the latent states.

\begin{table}[ht]
\centering
\caption{Posterior means and standard deviations of the GIHMM parameters, together with the corresponding true values.}
\label{tab:2}
\begin{tabular}{lrr}
\hline
Parameter & Posterior mean (sd) & True value \\
\hline
$\mu^*_{1,1}$ & 35.00 (0.05) & 35.00 \\
$\mu^*_{1,2}$ & 35.01 (0.05) & 35.00 \\
$\mu^*_{2,1}$ & 40.02 (0.05) & 40.00 \\
$\mu^*_{2,2}$ & 40.01 (0.05) & 40.00 \\
$\sigma^*_{1,1,1}$ & 0.96 (0.06) & 1.00 \\
$\sigma^*_{1,2,2}$ & 1.11 (0.08) & 1.00 \\
$\sigma^*_{2,1,1}$ & 0.94 (0.06) & 1.00 \\
$\sigma^*_{2,2,2}$ & 0.97 (0.06) & 1.00 \\
\hline
\end{tabular}
\end{table}

Table \ref{tab:2} shows that the gaussian parameters are accurately recovered. In particular, the posterior means of the state-specific means are almost identical to the corresponding true values, with discrepancies of at most 0.02. The variance-covariance parameters are also estimated with good accuracy, although a slight upward bias is observed for $\sigma^*_{1,2,2}$ and a slight downward bias for $\sigma^*_{1,1,1}$ and $\sigma^*_{2,1,1}$. Nevertheless, these differences are small relative to the posterior uncertainty. Overall, the low posterior standard deviations indicate that these parameters are estimated precisely, confirming that the MCMC proposed for the estimation of the GIHMM is able to recover the emission distribution, that is, the distribution of the observable variables, effectively. Regarding the other parameters, Table \ref{tab:3} reports the posterior means, standard deviations, and the corresponding true values.

\begin{table}[ht]
\centering
\caption{Posterior means and standard deviations of the GIHMM parameters, together with the corresponding true values.}
\label{tab:3}
\begin{tabular}{lrr}
\hline
Parameter & Posterior mean (sd) & True value \\
\hline
$\omega^*_{1,1,1}$ & 0.37 (0.03) & 0.31 \\
$\omega^*_{1,2,1}$ & 0.74 (0.03) & 0.79\\
$\omega^*_{2,1,1}$ & 0.80 (0.03) & 0.69 \\
$\omega^*_{2,2,1}$ & 0.21 (0.03) & 0.21 \\
\hline
\end{tabular}
\end{table}

Looking at Table \ref{tab:3}, the posterior estimates closely recover the data-generating parameters, indicating that the proposed MCMC for the GIHMM provides accurate inference. 
The transition probabilities are generally well estimated, with only a moderate bias for $\omega^*_{1,1,1}$ and $\omega^*_{2,1,1}$. In addition, we can notice that the posterior uncertainty is low, as reflected by the small posterior standard deviations.

\subsection{Synthetic dataset 2: The Homogeneous Case}
This simulation is designed to assess whether the proposed GIHMM can recover a standard infinite HMM, that is, when no temporal heterogeneity is present in the transition dynamics. In particular, in the second simulated dataset we generate observations from a model with a homogeneous transition probability matrix. We consider the same number of units ($N=30$), the same number of time points ($T=30$), and the same true number of latent states, as in Section \ref{sec:sim1}. Moreover, the prior hyperparameters values and the MCMC algorithm are identical to those adopted in Section \ref{sec:sim1}. The transition probability matrix is fixed over time and is given by
\[
\begin{bmatrix}
0.31 & 0.69\\
0.79 & 0.21
\end{bmatrix}.
\]

Also for the vector means, we consider the same structure previously defined, where we assume $[35,35]'$ and $[40,40]'$, while for the variance-covariance matrices we use an identity matrix. 

\begin{figure}[!htbp]
\centering
\includegraphics[width=0.7\textwidth]{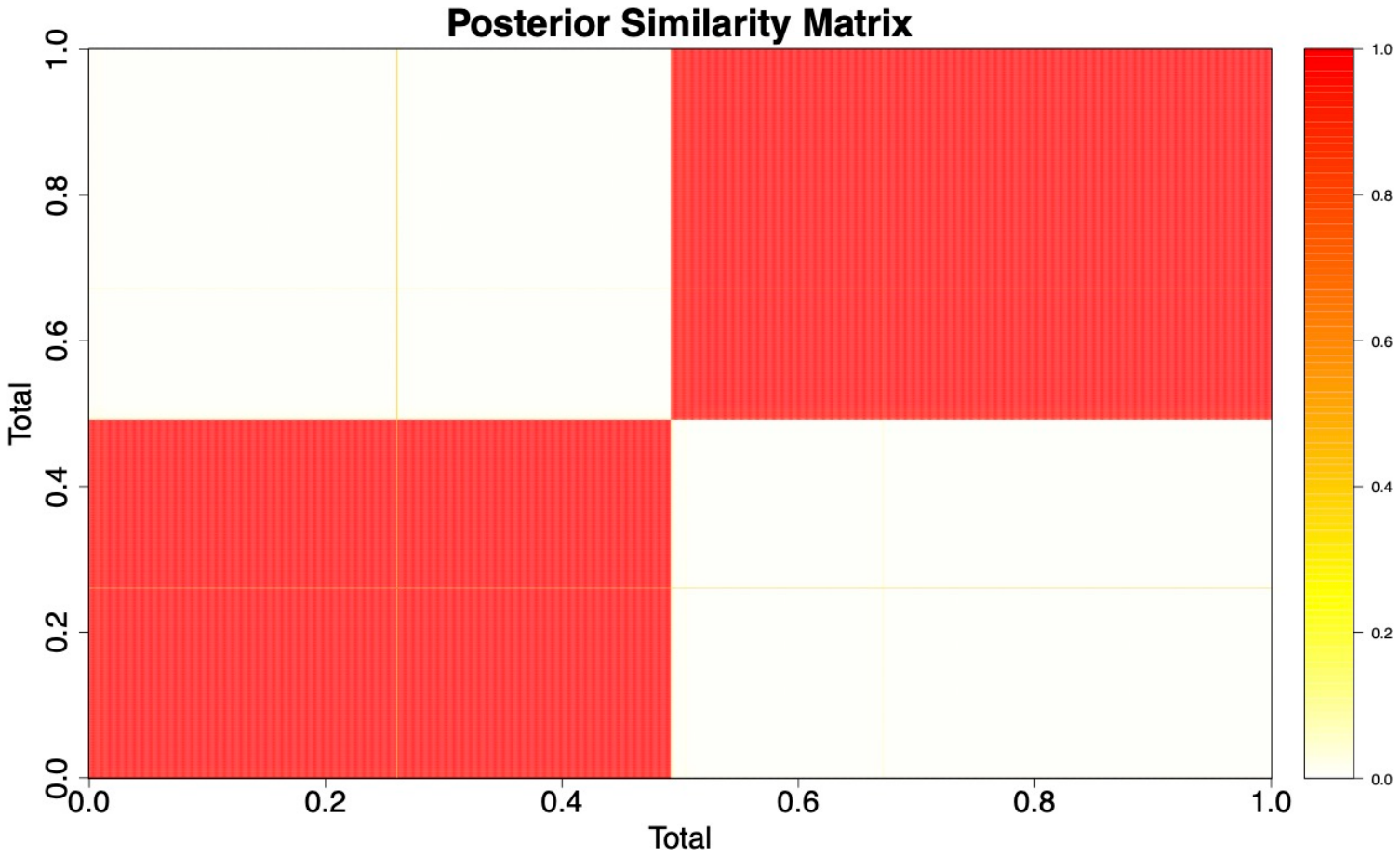}
\caption{Synthetic dataset 2. Posterior similarity matrix for $\boldsymbol{V}$ of the GIHMM. Total, on the $x$- and $y$- axes, means that all the $V_{i,t}$ are considered.}
\label{fig4}
\end{figure}

\begin{figure}[!htbp]
\centering
\includegraphics[width=0.7\textwidth]{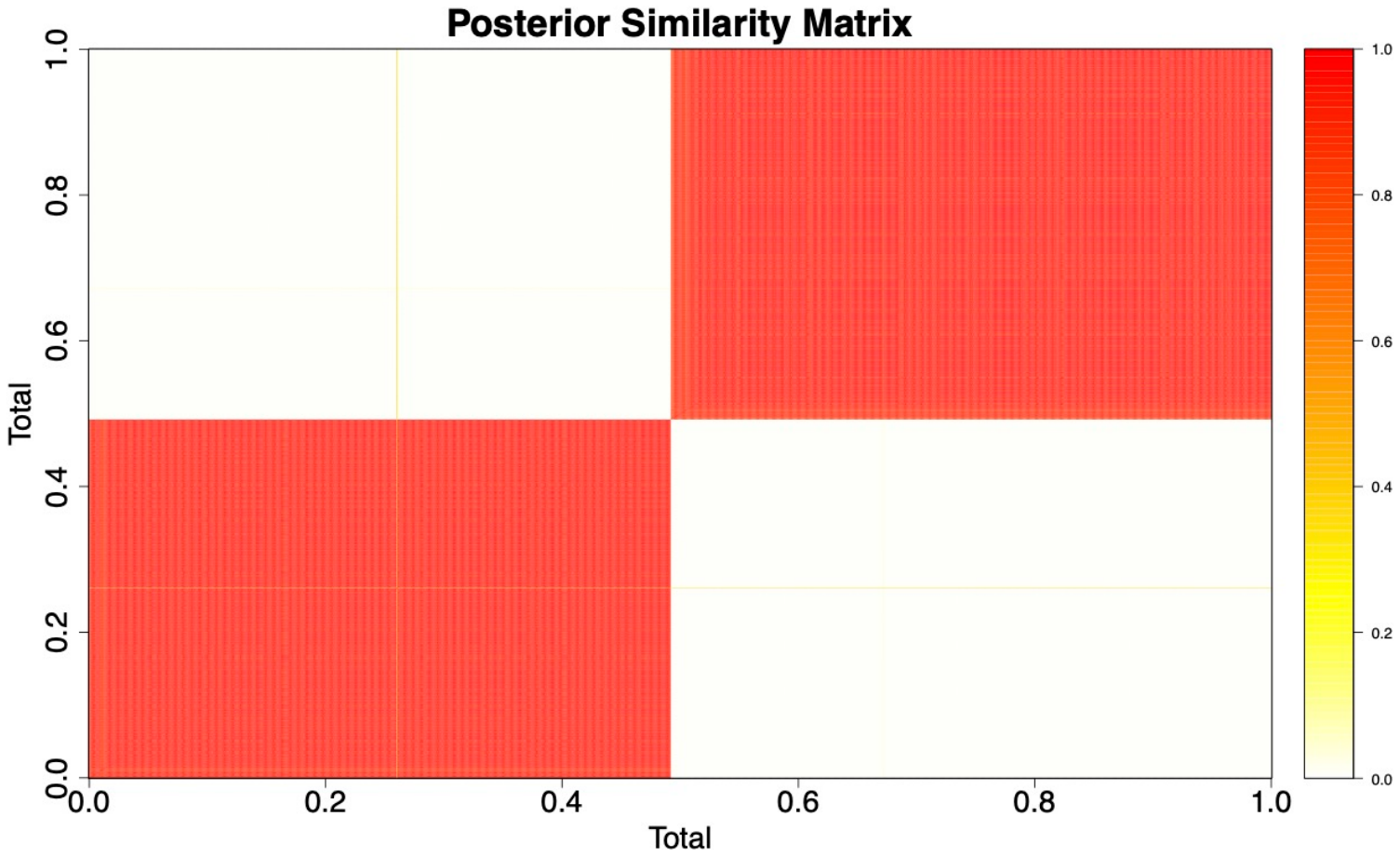}
\caption{Synthetic dataset 2. Posterior similarity matrix for $\boldsymbol{V}$ of the infinite HMM. Total, on the $x$- and $y$- axes, means that all the $V_{i,t}$ are considered.}
\label{fig5}
\end{figure}

We fit both the GIHMM and the infinite HMM to the synthetic dataset and compare their performance. As in the previous scenario, the final partition of the latent variables is obtained by minimizing the VI loss. In this setting, the GIHMM correctly identifies a single cluster for each transition row $j=1,2$, indicating that no temporal variation in the transition structure is detected. This result is consistent with the data-generating process, which assumes a homogeneous transition probability matrix over time. Regarding the estimated parameters and the $\boldsymbol{V}$ latent structure, the GIHMM correctly identifies two latent states, corresponding to the true number of states used to generate the data. Similarly, the infinite HMM is also able to recover the correct number of latent states. Figures \ref{fig4} and \ref{fig5} show the posterior similarity matrices obtained from the GIHMM and the infinite HMM, respectively.

Notice that, the plots in Figures \ref{fig4} and \ref{fig5} are identical. This means that the GIHMM is able to recover the underlying homogeneous structure by avoiding unnecessary additional clusters. We further evaluate the performance of the two approaches by computing the classification accuracy for both the GIHMM and the infinite HMM. Table \ref{tab:4} reports the classification accuracy associated with the latent structures of the two models. In particular, for the GIHMM, we consider both latent processes, $\boldsymbol{V}$ and $\boldsymbol{Z}$, while for the infinite HMM we evaluate the accuracy only for the latent state sequence $\boldsymbol{V}$.

\begin{table}[ht]
\centering
\caption{Classification accuracy for the GIHMM and the infinite HMM.}
\label{tab:4}
\begin{tabular}{llr}
\hline
GIHMM & & \\
\hline
& Latent structure & Accuracy $(\%)$ \\
\hline
& $\boldsymbol{V}$ & 99.89 \\
& $\boldsymbol{Z}$ & 100.00 \\
\hline
Infinite HMM & & \\
\hline
& Latent structure & Accuracy $(\%)$ \\
\hline
& $\boldsymbol{V}$ & 99.89 \\
\hline
\end{tabular}
\end{table}

Table \ref{tab:4} shows that both models achieve excellent classification accuracy in recovering the latent state sequence. In particular, the GIHMM correctly identifies the latent process $\boldsymbol{V}$ with an accuracy of $99.89\%$, while the additional latent structure $\boldsymbol{Z}$, which governs the clustering of the transition probability rows, is recovered perfectly with an accuracy of $100.00\%$. Similarly, the infinite HMM achieves an accuracy of $99.89\%$ for the latent states $\boldsymbol{V}$.

Regarding the estimated parameters, Table \ref{tab:5} reports the posterior means and standard deviations of the GIHMM and the infinite HMM, together with the corresponding true values. 

\begin{table}[ht]
\centering
\caption{Posterior means and standard deviations of the GIHMM and infinite HMM parameters, together with the corresponding true values.}
\label{tab:5}
\begin{tabular}{lrrr}
\hline
& \multicolumn{1}{c}{GIHMM} & \multicolumn{1}{c}{Infinite HMM} & \\
\cline{2-3}
Parameter & Posterior mean (sd) & Posterior mean (sd) & True value \\
\hline
$\mu^*_{1,1}$ & 34.98 (0.05) & 34.98 (0.05) & 35.00 \\
$\mu^*_{1,2}$ & 35.05 (0.05) & 35.05 (0.05) & 35.00 \\
$\mu^*_{2,1}$ & 40.04 (0.04) & 40.04 (0.04) & 40.00 \\
$\mu^*_{2,2}$ & 39.96 (0.05) & 39.96 (0.05) & 40.00 \\
$\sigma^*_{1,1,1}$ & 1.04 (0.08) & 1.04 (0.07) & 1.00 \\
$\sigma^*_{1,2,2}$ & 0.99 (0.08) & 0.98 (0.07) & 1.00 \\
$\sigma^*_{2,1,1}$ & 0.86 (0.07) & 0.86 (0.06) & 1.00 \\
$\sigma^*_{2,2,2}$ & 1.09 (0.08) & 1.08 (0.07) & 1.00 \\
$\omega^*_{1,1,1}$ & 0.30 (0.02) & 0.30 (0.02) & 0.31 \\
$\omega^*_{1,2,1}$ & 0.68 (0.02) & 0.68 (0.02) & 0.79 \\
\hline
\end{tabular}
\end{table}

Table \ref{tab:5} shows that the two models provide very similar estimates for the Gaussian parameters, with posterior means close to the data-generating values. In particular, the state-specific means $\mu^*_{1,1}$, $\mu^*_{1,2}$, $\mu^*_{2,1}$, and $\mu^*_{2,2}$ are accurately recovered by both approaches. The variance-covariance parameters are also well estimated, although slightly larger discrepancies are observed for $\sigma^*_{2,1,1}$, whose posterior mean is lower than the true value. Nevertheless, these differences remain limited.
The transition probabilities are well recovered by both models, with posterior means very close to the true values, with the exception of $\omega^*_{1,2,1}$, which is slightly underestimated. Since this scenario is generated under a homogeneous transition structure, the comparable performance of the GIHMM and the infinite HMM is expected. 

Overall, these results show that the proposed GIHMM is able to adapt to a homogeneous setting, maintaining estimation accuracy comparable to the infinite HMM.

\subsection{Longitudinal Multivariate Analysis of World Development Indicators}

In order to illustrate the proposed GIHMM, we consider the dataset analyzed in \cite{pennoni2024variable}, which contains 25 socioeconomic indicators from the World Development Indicators (WDI) collected by the World Bank and the UNESCO Institute for Statistics. Additional details on the dataset and the indicators are available at \url{https://databank.worldbank.org/data/source/world-development-indicators}. 

Since our model assumes Gaussian responses, we focus on a subset of 8 continuous indicators among the original 25 variables. In particular, we consider: life expectancy, infant mortality rate, gross national expenditure, GDP per capita, unemployment rate, internet usage, trade, and fertility rate. These indicators provide a comprehensive description of countries demographic, economic, social, and technological characteristics.
Furthermore, the original dataset contains several missing observations. To obtain a complete dataset, we retain only countries with complete observations for all selected indicators and all considered time points. After this preprocessing step, the final dataset consists of 131 countries observed over 18 years.

We set $L=J=10$ and $K=5$, while for the hyperparameters we impose $\beta = 1, \alpha = 1, \beta_0 = 1, s = 12, \boldsymbol{S} = s \times \boldsymbol{I}, \boldsymbol{m} = \boldsymbol{0}, \boldsymbol{M} = 100 \times \boldsymbol{I}$. The MCMC algorithm is run for 20,000 MCMC iterations, with the first $50\%$ discarded as burn-in. No thinning is applied. The final partition of the latent variables is obtained by minimizing VI loss. 

\begin{figure}[!htbp]
\centering
\includegraphics[width=0.7\textwidth]{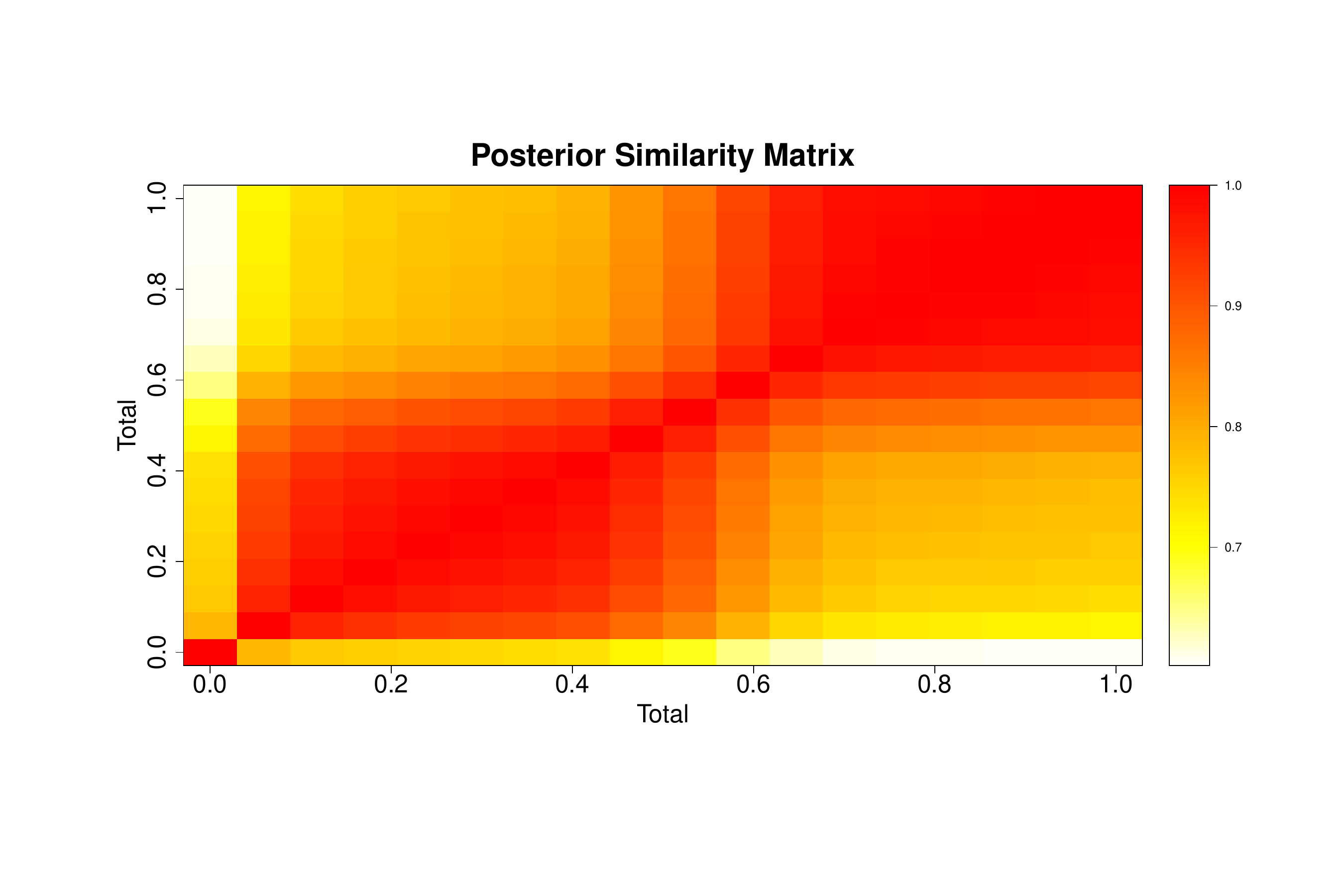}
\caption{Real dataset 1. Posterior similarity matrix for the first country in $\boldsymbol{V}$ of the GIHMM. Total, on the $x$- and $y$- axes, means that all the $V_{1,t}$ are considered.}
\label{fig6}
\end{figure}

The GIHMM identifies 5 latent states, while the latent variables in $\boldsymbol{Z}$ identify only a single transition probability matrix that does not change over time. The estimated number of latent states is broadly consistent with the results reported in \cite{pennoni2024variable}, who identified 6 latent states. The slight discrepancy is likely due to differences in the datasets used for estimation. The analysis in \cite{pennoni2024variable} is conducted on the complete dataset by employing a missing data imputation procedure, whereas our analysis is restricted to countries with complete observations. Consequently, the reduced sample size may explain why our model identifies one fewer latent state. 
The posterior similarity matrix for the latent variables in $\boldsymbol{V}$ of the GIHMM is shown in Figure \ref{fig6}.

Table \ref{tab:6} reports the posterior means and standard deviations of the state-specific means and variances associated with the GDP per capita variable estimated under the GIHMM.

\begin{table}[ht]
\centering
\caption{Posterior means and standard deviations of the means and variances of the GIHMM related to the variable GDP per capita.}
\label{tab:6}
\begin{tabular}{lrr}
\hline
Parameter & Posterior mean & Posterior sd \\
\hline
$\mu^*_{1}$ & 11.32 & 0.32  \\
$\mu^*_{2}$ & 12.08 & 0.08  \\
$\mu^*_{3}$ & 14.02 & 0.04  \\
$\mu^*_{4}$ & 13.21 & 0.16  \\
$\mu^*_{5}$ & 10.39 & 0.07   \\
$\sigma^*_{1}$ & 6.75 & 0.79  \\
$\sigma^*_{2}$ & 0.64 & 0.09  \\
$\sigma^*_{3}$ & 0.74 & 0.05  \\
$\sigma^*_{4}$ & 3.27 & 0.26  \\
$\sigma^*_{5}$ & 1.81 & 0.11  \\
\hline
\end{tabular}
\end{table}

Notice that the estimated means can be ordered from the lowest to the highest values, allowing an interpretation of the identified latent states. For example, in the case of GDP per capita, the states can be interpreted as representing groups of countries with different levels of economic development, ranging from countries with lower GDP per capita to countries with higher GDP per capita. Regarding the estimated transition probability matrix, this exhibits a strong persistence in the latent dynamics, with the highest probabilities concentrated along the main diagonal. In particular, the diagonal elements indicate that, once a country enters a given latent state, it is highly likely to remain in that state at the following time point. The results are reported in Table \ref{tab:7}.

\begin{figure}[!htbp]
\centering
\includegraphics[width=0.7\textwidth]{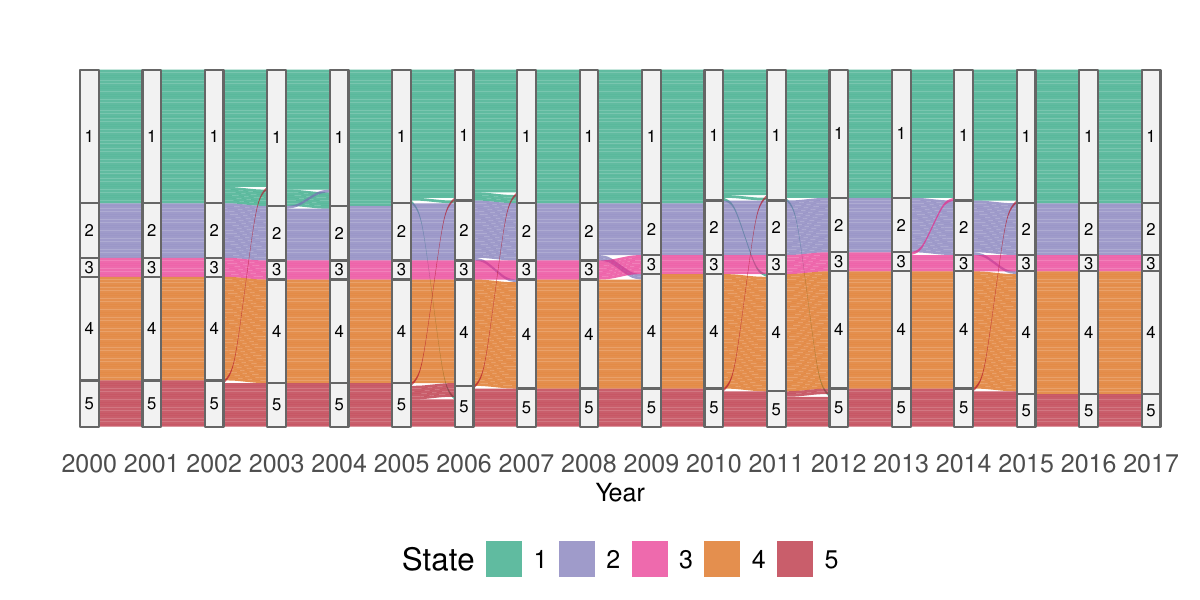}
\caption{Real Dataset 1. Evolution of latent variables $V_{i,t}$ across countries over time.}
\label{fig8}
\end{figure}

\begin{table}[ht]
\centering
\caption{Posterior means and standard deviations of the diagonal elements of the transition probability matrix of the GIHMM.}
\label{tab:7}
\begin{tabular}{lrr}
\hline
Parameter & Posterior mean & Posterior sd \\
\hline
$\omega^*_{1,1,1}$ & 0.92 & 0.04  \\
$\omega^*_{1,2,2}$ & 0.98 & 0.01  \\
$\omega^*_{1,3,3}$ & 0.99 & 0.00  \\
$\omega^*_{1,4,4}$ & 0.97 & 0.02  \\
$\omega^*_{1,5,5}$ & 0.98 & 0.01  \\
\hline
\end{tabular}
\end{table}
Also in this case, the results are consistent with those reported in \cite{pennoni2024variable}, who identified the same structure for the transition probability matrix. Figure \ref{fig8} represents the evolutions of the latent variables over time.

 \subsection{Unfolding Epochs in State-Level Production Data}
 
The second application is based on the \textbf{produc} dataset \citep{baltagi2008econometric} from the \textbf{Ecdat} R package. It contains annual observations for the 48 contiguous U.S. states over the period 1970-1986. The dataset includes information on economic production, input factors, and labour market characteristics. In this analysis, we consider a subset of 7 variables. These are: private capital stock per capita, highway infrastructure per capita, water infrastructure per capita, utility infrastructure per capita, private consumption per capita, gross state product, and unemployment rate. In particular, as a response variable, we consider the differences over the years, that is $Y_{i,t} = X_{i,t} - X_{i,t-1}$, where $X_{i,t}$ are the original data. 

\begin{figure}[!htbp]
\centering
\includegraphics[width=0.7\textwidth]{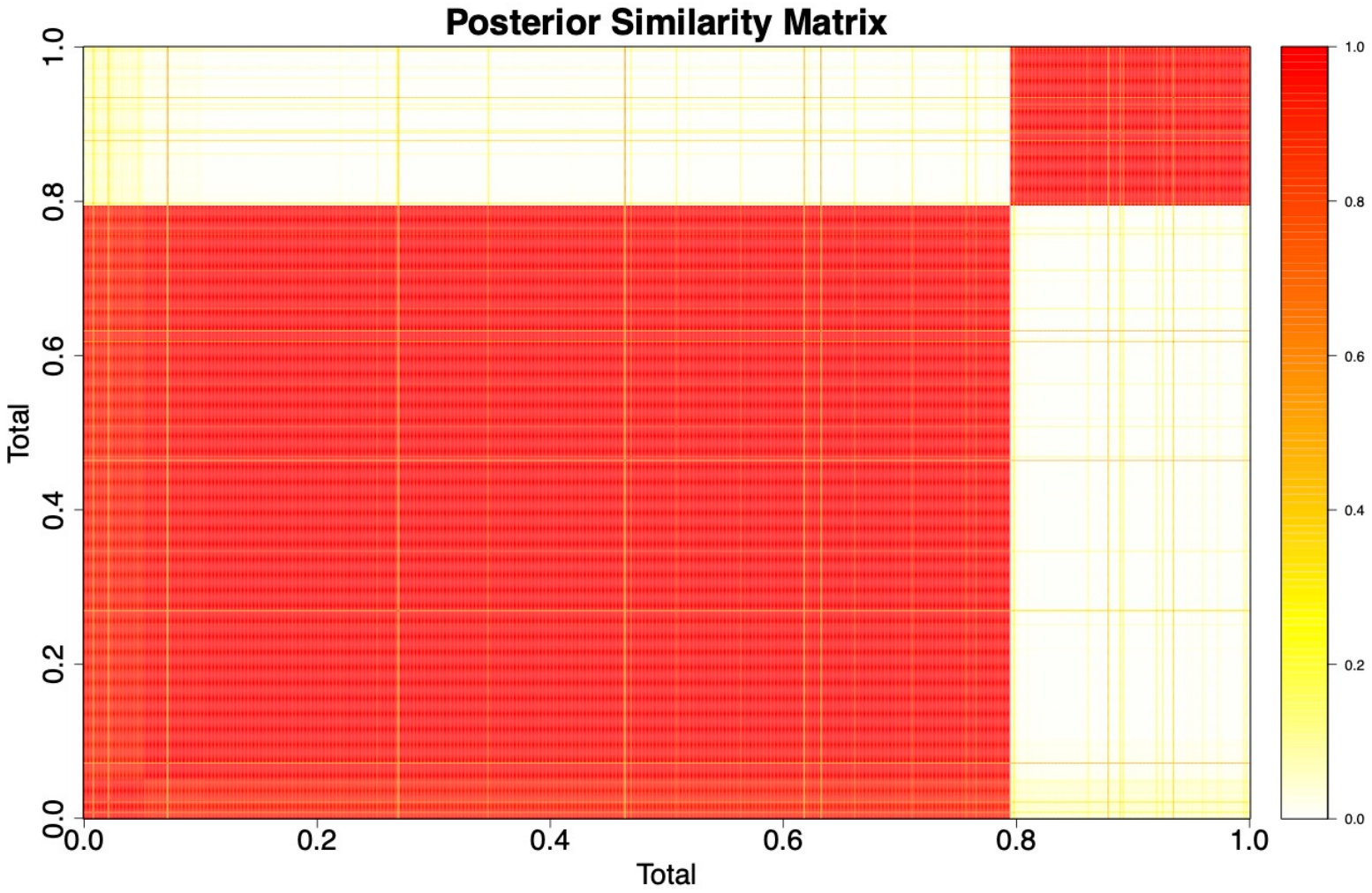}
\caption{\textbf{Produc} dataset. Posterior similarity matrix for all the U.S. states in $\boldsymbol{V}$ of the GIHMM. Total, on the $x$- and $y$- axes, means that all the $V_{i,t}$ are considered.}
\label{fig7}
\end{figure}

We fit the GIHMM proposed and the model identifies 2 latent states, while the latent variables in $\boldsymbol{Z}$ identifies mainly a single transition probability matrix, with 2 different changes, maintaining invariant the transition over the other time periods. 
The posterior similarity matrix for the latent variables in $\boldsymbol{V}$ of the GIHMM is shown in Figure \ref{fig7}.

The estimated parameters are reported in Table \ref{tab:8}. In particular, we report the estimated means and variances associated with the variable private capital stock per capita, together with the estimated transition probabilities. The results show that the transition structure is mostly characterized by a single configuration, with the exception of two time points where a different transition mechanism is identified. The dominant transition dynamics are described by $\omega^*_{1,1,1}$ and $\omega^*_{1,2,2}$, which correspond to the most frequently occurring transition probability matrix over time. The alternative transition structure is identified through $\omega^*_{2,1,1}$ and $\omega^*_{2,2,2}$, capturing the temporal variation detected by the GIHMM.
 
 \begin{table}[ht]
\centering
\caption{Posterior means and standard deviations of the means and variances of the GIHMM related to the variable private capital stock per capita. In addition, also the transition probabilities are reported.}
\label{tab:8}
\begin{tabular}{lrr}
\hline
Parameter & Posterior mean & Posterior sd \\
\hline
$\mu^*_{1,1}$ & 0.27 & 0.01  \\
$\mu^*_{2,1}$ & 0.93 & 0.09  \\
$\sigma^*_{1,1,1}$ & 0.10 & 0.01  \\
$\sigma^*_{2,1,1}$ & 1.32 & 0.15  \\
$\omega^*_{1,1,1}$ & 0.97 & 0.01  \\
$\omega^*_{1,2,2}$ & 0.95 & 0.03  \\
$\omega^*_{2,1,1}$ & 0.79 & 0.14  \\
$\omega^*_{2,2,2}$ & 0.92 & 0.07  \\
\hline
\end{tabular}
\end{table}

\begin{figure}[!htbp]
\centering
\includegraphics[width=0.7\textwidth]{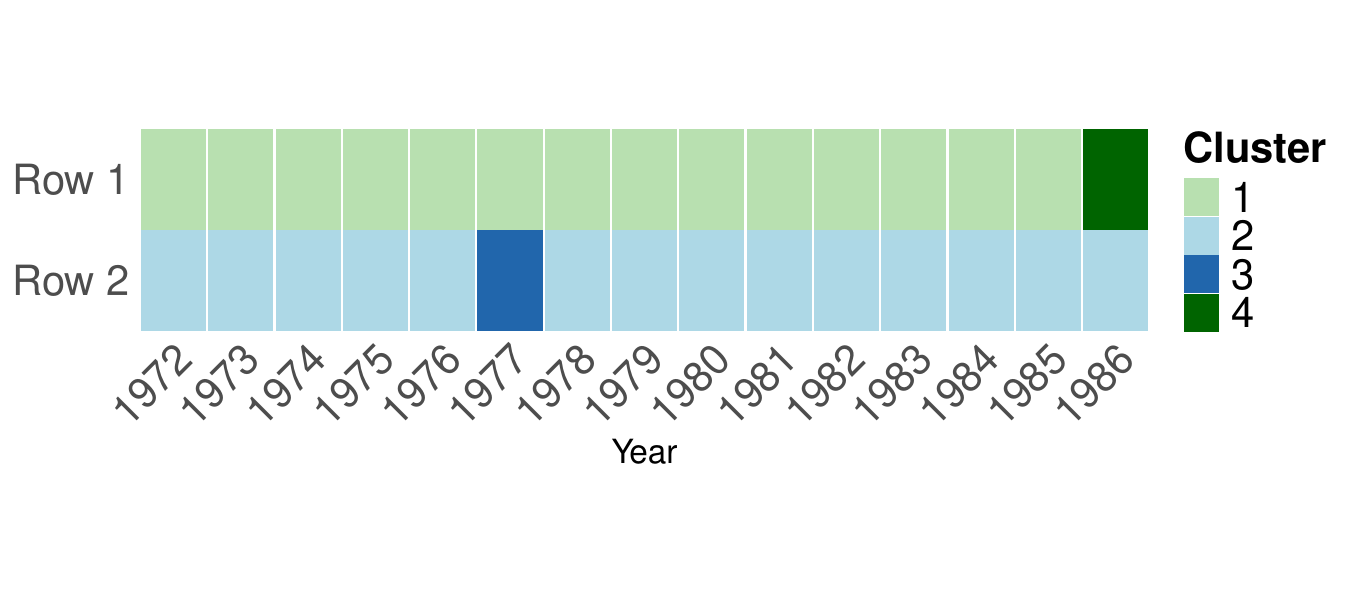}
\caption{Real Dataset 2. Evolution of latent variables $Z_{j,t}$ across years.}
\label{fig10}
\end{figure}

Table \ref{tab:8} shows that the estimated transition matrix is characterized by a strong persistence in the latent dynamics, with a high probability of remaining in the same state over time. Furthermore, the GIHMM identifies 2 time points (years), $1977$ and $1986$, where the transition structure differs from the main configuration. In $1977$, the second row of the transition matrix changes, resulting in a lower probability of remaining in the same state compared with the predominant transition structure. Similarly, in $1986$, a different first row is identified, which is also characterized by a reduced persistence probability relative to the main transition mechanism. These results suggest that, although the latent process is mostly stable over time, the model is able to detect specific periods characterized by changes in the transition dynamics. A graphical representation is provided in Figure \ref{fig10}.

\section{Conclusions} \label{sec:6}

In this paper we propose a generalized hidden Markov model (HMM) that relaxes the homogeneity assumption implicit in the infinite HMM of \citet{Beal_2002,Teh_2006}. The proposed model is based on the hidden hierarchical Dirichlet process proposed by \citet{Lijoi_2023}. In particular, the generalized infinite HMM combines the standard (time-homogeneous) infinite HMM as a special case, yet it is flexible enough to detect changes and change points in the transition dynamics, allowing the rows of the transition probability matrix to be clustered in a suitable way.

Bayesian estimation is performed by using a Markov chain Monte Carlo algorithm. In particular, we consider a Metropolis-within-Gibbs algorithm, which is based on the finite approximation of the Dirichlet process. In the proposed algorithm, all model components are estimated jointly within a single procedure, avoiding the comparison of a potentially prohibitive number of parametric specifications.

The comparison of the classical infinite HMM with the proposed generalized infinite HMM shows that the latter can accurately recover the underlying structure in both homogeneous and heterogeneous settings. In contrast, when the classical infinite HMM is applied to heterogeneous data, it tends to overestimate the number of components. Finally, we illustrate the flexibility of the proposed model through two real-data applications.

The proposed approach also has some limitations. The prior distributions treat time points as exchangeable within each departure state, thus ignoring the temporal ordering and not favouring contiguous regimes, and it clusters the transition kernels independently across departure states, so that a break affecting the whole transition matrix must be detected row by row. Both issues could be addressed by a Markovian or sticky allocation of the variables $Z_{j,t}$ and by sharing these variables across departure states. More generally, since the hidden hierarchical Dirichlet process clusters populations of transitions irrespective of how these are defined, time points could be replaced or combined with discrete covariates, while continuous covariates could enter through covariate-dependent stick-breaking weights for $Q_j$. Finally, since the prior on the transition dynamics does not depend on the emission model, the approach extends directly to discrete, mixed, or other non-Gaussian responses. These extensions are left for future investigations. 

\subsubsection*{Declaration of generative AI in the manuscript preparation process}
During the preparation of this article ChatGPT was used in order to detect any spelling and grammar mistakes. After using this tool, the authors reviewed and edited the content as needed and take full responsibility of the content of the published article.

\section*{Appendix A}

In this section, we show how to derive the full conditional distribution discussed in Section \ref{sec:4}. First of all, notice that $p(\boldsymbol{y} \lvert \boldsymbol{\theta}^*, \boldsymbol{v}) p(\boldsymbol{v} \lvert \boldsymbol{z}, \boldsymbol{\omega}^*) p(\boldsymbol{z} \lvert \boldsymbol{\pi})$ may be written considering suitable indicator functions as a product of three different components:
\begin{equation}
\begin{split}
I_1 = & \prod_{i=1}^N\prod_{l=1}^L \left[ \omega_{l} \phi(\boldsymbol{y}_{i,1}; \boldsymbol{\mu}^*_l, \boldsymbol{\Sigma}^*_{l}) \right]^{\mathds{1}(v_{i,1} =l)} \\
I_2 = & \prod_{t=2}^T \prod_{j=1}^{J} \prod_{k=1}^K \pi_{k,j}^{\mathds{1}(z_{j,t} =k)} \prod_{l=1}^L \prod_{i=1}^N \omega_{k,j,l}^{*\mathds{1}(z_{j,t} =k)\mathds{1}(v_{i,t-1} =j,v_{i,t} =l)} \\
I_3 = & \prod_{t=2}^T \prod_{i=1}^N \prod_{l=1}^L \phi(\boldsymbol{y}_{i,t}; \boldsymbol{\mu}^*_l, \boldsymbol{\Sigma}^*_{l})^{\mathds{1}(v_{i,t} =l)},
\end{split}
\label{eq:prod_lik}
\end{equation}
where $\phi(\cdot)$ denotes the multivariate normal density, and, as already defined in Section \ref{sec:exthmm}, the initial probability $\omega_{l}$ is assumed to be known. Based on Equations \eqref{eq:prod_lik}, we have
\begin{equation*}
\begin{split}
p(Z_{j,t} = k \lvert \cdots) & \propto \pi_{k,j} \prod_{l=1}^L \prod_{i=1}^N \omega_{k,j,l}^{*\mathds{1}(v_{i,t-1}=j, v_{i,t} = l)}, 
\end{split}
\end{equation*}
where $j=1,\ldots,J$, $k=1,\ldots, K$, and $t=2,\ldots,T$. In addition, combining Equations \eqref{eq:prod_lik} and \eqref{eq:post}, we have
\begin{equation}
\begin{split}
p(\pi_{j} \lvert \cdots) & \propto \prod_{k=1}^K \pi_{k,j}^{\alpha_k/K -1} \times \prod_{k=1}^K \prod_{t=2}^T \pi_{k,j}^{\mathds{1}(z_{j,t}=k)} \\
& = \prod_{k=1}^K \pi_{k,j}^{\alpha_k/K + \sum_{t=2}^T \mathds{1}(z_{j,t}=k)-1},
\end{split}
\end{equation}
from which we can recognize the Dirichlet kernel, $\mathcal{D}ir(\boldsymbol{\alpha}/K + \boldsymbol{m}_j)$, $j=1,\ldots,J$,
where 
$$\boldsymbol{m}_j = \left[ \sum_{t=2}^T \mathds{1}(z_{j,t} =1), \ldots, \sum_{t=2}^T \mathds{1}(z_{j,t} =K) \right]'.$$
Furthermore, we have
\begin{equation}
\begin{split}
p(\boldsymbol{\omega}^*_{k,j} \lvert \cdots) & \propto \prod_{l=1}^L \omega_{k,j,l}^{*\beta \omega_{0,l}-1} \times \prod_{l=1}^L \prod_{t=2}^T \prod_{i=1}^N \omega_{k,j,l}^{*\mathds{1}(z_{j,t}=k) \mathds{1}(v_{i,t-1}=j, v_{i,t}=l)} \\
& = \prod_{l=1}^L \omega_{kj,l}^{*\beta \omega_{0,l} + \sum_{t=2}^T \mathds{1}(z_{j,t}=k) \sum_{i=1}^N \mathds{1}(v_{i,t-1}=j, v_{i,t}=l)-1},
\end{split}
\end{equation}
from which we can recognize the Dirichlet kernel, $\mathcal{D}ir(\beta\boldsymbol{\omega}_0 + \boldsymbol{n}_{k,j})$, $j=1,\ldots,J, k=1,\ldots,K$, where 
$$\boldsymbol{n}_{k,j} = \left[ \sum_{t=2}^T \mathds{1}(z_{j,t} =k) \sum_{i=1}^N \mathds{1}(v_{i,t-1} = j, v_{i,t} = 1), \ldots, \sum_{t=2}^T \mathds{1}(z_{j,t} =k) \sum_{i=1}^N \mathds{1}(v_{i,t-1} = j, v_{i,t} = L) \right]'.$$ 

Regarding Equation \eqref{eq:ful_v}, this is simply obtained considering the elements that contain $V_{i,t}$, while Equations \eqref{eq:11} and \eqref{eq:12} are standard full conditionals for Gaussian and Inverse-Wishart cases; see \citet{tancini2024comparison}. 

\bibliographystyle{apacite}
\bibliography{reference}
\end{document}